\documentclass[11pt, a4paper]{article}
\usepackage{jheparxiv}
\usepackage[utf8]{inputenc}
\usepackage{amsmath}
\usepackage{comment}
\newcommand{\be}{\begin{equation}}
\newcommand{\ee}{\end{equation}}
\usepackage{amsmath,bm}
\usepackage[all]{xy}
\usepackage{tikz-cd}
\usepackage{dsfont}
\usepackage{bm}
\usepackage{amsfonts}
\usepackage{appendix}
\usepackage{amssymb}
\usepackage{hyphenat}
\usepackage{latexsym}
\usepackage{float}
\usepackage{mathrsfs}
\usepackage{braket}	
\usepackage{graphicx}
\usepackage{color}
\usepackage{xcolor}
\usepackage{slashed}
\usepackage{mathtools}
\usepackage{breqn}
\usepackage[all]{xy}
\usepackage{tikz-feynman} 
\usepackage{dsfont}
\usepackage{cancel}
\usepackage{amsfonts}
\usepackage{amssymb}
\usepackage{amsmath,amssymb}
\usepackage{booktabs}
\usepackage{array}
\usepackage{tabularx}
\usepackage{hyperref}
\usepackage{pifont}
\usepackage{bm}
\usepackage{float}
\usepackage{amsmath}
\usetikzlibrary{arrows.meta}
\newcolumntype{Y}{>{\raggedright\arraybackslash}X}
\newcommand{\adot}{\dot{\alpha}}

\begin{document}
\newcommand{\RNum}[1]{\uppercase\expandafter{\romannumeral #1\relax}}
\tikzfeynmanset{compat=1.1.0}
\subheader{\hfill \texttt{}}
\newcommand{\com}[1]{ \textcolor{red}{(#1)}}
\newcommand{\blue}[1]{ \textcolor{blue}{#1}}

\title{4d CFT Correlators from Ambitwistors}
\author[a,b]{Mariana Carrillo González,}
\author[a]{and Théo Keseman}
\affiliation[a]{Abdus Salam Centre for Theoretical Physics\\
Imperial College London, SW7 2AZ, United Kingdom}
\affiliation[b]{School of Physics \& Astronomy\\
University of Southampton, SO17 1BJ, United Kingdom}
\emailAdd{m.carrillo-gonzalez@imperial.ac.uk}
\emailAdd{theo.keseman17@imperial.ac.uk}
\abstract{
 We develop an ambitwistor-space formulation of conserved spinning correlators in four-dimensional CFTs. We show that conformal symmetry and conservation are trivialised by the ambitwistor Penrose transform, and construct the two- and three-point ambitwistor correlators explicitly. We further propose a simple formula for projecting onto parity-even and parity-odd sectors, and verify it in several non-trivial examples. The resulting correlators furnish a natural basis compatible with a boundary double copy for arbitrary spin. At three points, we show that they localise on degree-three curves in a two-twistor representation. We also explain how the formalism incorporates boundary propagators for unitary operators of integer conformal dimension, and extend it to conformally coupled scalars.}

\maketitle

\newpage
\section{Introduction}
The study of scattering amplitudes has revealed that physical observables often admit descriptions which are far simpler than those suggested by a direct Feynman-diagram expansion. In flat space, spinor-helicity variables make Lorentz invariance and little-group covariance manifest, leading to compact expressions such as the Parke--Taylor formula for MHV amplitudes \cite{Parke:1986gb}. The remarkable simplicity of these formulae is not accidental: in twistor space, the kinematical constraints of massless scattering are geometrised, and amplitudes acquire a natural interpretation in terms of holomorphic data \cite{Witten:2003nn, Berkovits:2004hg, smatrix, Mason:2009sa}.

A parallel simplification is highly desirable in curved spacetimes, in particular for boundary correlators in Anti-de Sitter (AdS) and wavefunction coefficients in de Sitter (dS). In the conventional embedding-space approach, conformal symmetry is manifest, but the construction of spinning correlators remains technically involved. The tensor structures must be built from conformally invariant building blocks and then constrained by conservation equations, permutation symmetries, Ward identities, and, when appropriate, parity. These constraints become increasingly difficult to solve as the spin and the number of points grow. In particular, the understanding of massless spinning fields in $(A)dS_5$, or equivalently conserved operators on the boundary, constitute a physically relevant case that will be under consideration here. Incidentally, this is a non-trivial situation to consider, since in this case the representation shortens, and in practice, not all conformally invariant structures are allowed. It is therefore reasonable to wonder if the conservation condition can be trivialised. 

The aim of this paper is to show that many of these difficulties are substantially simplified by formulating four-dimensional CFT correlators in ambitwistor space, which is the space of complex null geodesics \cite{Mason:2005kn}. The idea to find a space that simplifies correlators of conformally invariant theories is not new, and many such languages already exist, for instance embedding space \cite{Costa:2011mg, Zhiboedov:2012bm} (and its bispinor formulation \cite{Binder:2020raz,Simmons-Duffin:2012juh}), Mellin space \cite{Penedones:2010ue}, momentum space \cite{Bzowski:2013sza,Coriano:2013jba,Baumann:2022jpr}, and analytic superspace \cite{Heslop:2025zeq}. For four-dimensional CFTs, we show that twistor variables  provide a natural home for conserved boundary correlators. The present paper is thus the natural generalisation of a program initiated in three dimensions (3d) where it was found that a twistor formulation existed and could drastically simplify conserved correlators \cite{Baumann:2024ttn, CarrilloGonzalez:2025qjk, Bala:2025gmz, Bala:2025jbh, Bala:2025qxr, Ansari:2025fvi}. In parallel, by trivialising conformal symmetry and the homogeneous Ward identity in momentum space, one is naturally led to the cosmological Grassmannian \cite{Arundine:2026fbr, De:2026shn, Huang:2026tsh, Arundine:2026myr}. These two constructions are intimately related. In 3d, the Grassmannian space arises as the Schwinger parametrisation of the twistors. Recently, it was found that a bi-symplectic Grassmannian can arise in the context of CFT$_4$ conserved correlators satisfying homogeneous Ward identities \cite{Bala:2026trw}.  In this paper, we formulate the ambitwistor construction and show that it can give rise to the correlators with inhomogeneous Ward identities. The precise correspondence between the ambitwistor and the Grassmannian sides in 4d is not yet clear, but we hope to clarify it in future work.

The central observation is that ambitwistor representatives trivialise several of the main difficulties associated with conserved spinning correlators.\footnote{More generally, as we will explain, our construction applies to all balanced mixed-symmetry twist-two correlators. In the scalar case, this corresponds in four dimensions to a field saturating the Breitenlohner-Freedman (BF) bound \cite{Breitenlohner:1982bm}. For symmetric traceless representations, the twist-two condition fixes the conformal dimension to the value appropriate for conserved operators.}. First, conformal invariance is built directly into the dependence on twistor and dual-twistor contractions. Second, bulk masslessness, equivalently boundary conservation, is encoded in the holomorphicity of the ambitwistor correlators and the projective properties of the Penrose transform. Third, the spin dependence is cleanly separated from the remaining functional dependence of the correlator. As a result, the problem of constructing conserved spinning correlators becomes almost as simple as in the scalar case. In this sense, the ambitwistor formulation reduces the construction of conserved spinning correlators to an algebraic problem, namely, finding representatives of the relevant cohomology classes on ambitwistor space.

\begin{raggedright}
A first goal of the paper is to construct two- and three-point functions in four-dimensional ambitwistor space. These correlators are fundamental because, in a conformal theory, two-point functions fix the normalization of operators, while three-point functions encode the OPE coefficients. Together with the spectrum, these quantities constitute the basic CFT data from which higher-point correlators can be reconstructed through the operator product expansion. Here, by working with suitable representatives of Čech cohomology classes, described locally by holomorphic functions on ambitwistor space, we obtain two- and three-point functions of conserved operators of arbitrary spin, without fixing their overall normalizations. 
\end{raggedright}

A second aim of this work is to make parity manifest in ambitwistor space. In three dimensions, for a fixed spin configuration, there is at most one parity-odd conserved structure, together with at most two parity-even ones. The four-dimensional case is richer, with multiple structures allowed in both parity sectors. We show that the ambitwistor construction provides a uniform way to organize this decomposition for arbitrary spin, and produces the full set of allowed tensor structures. This is especially suggestive because parity-odd structures are often associated with boundary anomalies \cite{Schreier:1971um,Giombi:2011rz,Coriano:2023hts}. Their appearance in a twistor or ambitwistor framework indicates that these methods are not limited to tree-level data, but can also capture a broader class of boundary observables, including loop-level effects.

There is an important new feature at three points. Unlike the simpler three-dimensional case, four-dimensional three-point functions in ambitwistor space admit a non-trivial cross-ratio. 
As a result, there is no obvious canonical choice of representatives for the relevant cohomology classes in ambitwistor space. This slightly obscures the direct relation between a natural boundary basis of ambitwistor correlators and a basis adapted to bulk interactions. Nevertheless, we will show that the boundary correlators associated with bulk Yang-Mills (YM) and General Relativity (GR) interactions can be written as a linear combination of the basis that is obtained from ambitwistor space.

As a consistency check of our formalism, we apply the Ward identity to a simple but non-trivial three-point function and extract the corresponding two-point function, which shows that our position-space expression captures the full correlator. This should be contrasted with the related Grassmannian approach, where one naturally obtains only the discontinuity of the correlator, which satisfies the homogeneous Ward identity. 

Perhaps surprisingly, ambitwistors can also be used beyond the conserved case. By considering the simplest deformation away from twist two, we show how any AdS boundary propagators for massive fields of arbitrary integer twist can be obtained directly from ambitwistor space. We also begin to investigate the half-integer twist case, and show in a simple example that such propagators can be recovered by choosing an appropriate Pochhammer contour. In particular, this allows us to obtain the (A)dS boundary propagator of a conformally coupled scalar.

The paper is organised as follows. In Section~\ref{sec:background}, we review the background material needed for the construction, including the embedding-space formalism for four-dimensional CFT correlators and ambitwistor space. In Section~\ref{sec:conserved-spinning-ambitwistor}, we construct conserved spinning correlators in ambitwistor space. We first discuss conformal invariance, show how conservation is implemented, analyse the action of parity, and develop the helicity-sector and derivative-spinning-factor construction. In Section~\ref{sec:three-point-functions}, we apply the formalism to two and three-point functions, beginning with the scalar case and then treating the general spinning case, which is obtained naturally as the application of twist-preserving weight-shifting operators which we express in ambitwistor space. We also show that, in a two-twistor representation, the scalar three-point function localises on a degree 3 curve. In Section~\ref{sec:properties}, we show that ambitwistor space naturally provides a basis of correlators adapted to a boundary double copy, and we verify the corresponding Ward identity. In Section~\ref{sec:beyond}, we explain how massive bulk fields are incorporated through a minimal deformation of the massless ansatz at 2-point and obtain the propagator of the conformally coupled scalar. 

Our conventions are collected in Appendix~\ref{app:conventions}. In Appendix~\ref{app:perform_penrose}, we explain the contour prescription and describe the numerical evaluation of the contour integrals, while Appendix~\ref{app:derivative-sector} contains examples of three-point functions evaluated in the derivative helicity sector. Finally, Appendix~\ref{app:dolbeault} presents the scalar three-point function in the Dolbeault, rather than \v Cech, representation, and Appendix~\ref{app:Pochhammer} reviews the Pochhammer contour and evaluates the contour integral relevant for the conformally coupled scalar.

\section{Background material}\label{sec:background}

As we will see, ambitwistor space provides a particularly simple description of correlation functions of twist-two operators. In this section, we fix notation and review the physical significance of these operators in four dimensions. We first recall their holographic/de Sitter interpretation, and then introduce the embedding-space and ambitwistor-space formalisms used throughout the paper. Readers familiar with these ingredients may safely skip this section.

\subsection{Twist-two operators and conservation} \label{sec:twist2}

In this paper, we will be concerned with four-dimensional conformal field theory (CFT) correlators. The bulk interpretation of these correlation functions is different for AdS or dS. 
In AdS, conserved currents and stress tensors are dual to massless bulk gauge fields and gravitons, with the shortening of the corresponding boundary conformal multiplets reflected as gauge symmetry in the bulk \cite{Skenderis:2002wp,DHoker:2002nbb}. More generally, in the standard quantisation, the leading near-boundary mode of a bulk field is fixed as the source, while the subleading mode determines the expectation value of the dual boundary operator. 

Meanwhile, in dS, the analogy holds at the level of representation theory, but the physical interpretation is different: expanding the wavefunction of the universe in late-time boundary fields defines wavefunction coefficients, rather than dual CFT expectation values. These wavefunction coefficients transform as conformal correlators on the future boundary, with massless gauge fields and gravitons corresponding to exceptional representations constrained by Ward identities from bulk gauge symmetry \cite{Maldacena:2002vr,Deser:2003gw}.

Local primary operators in a CFT are labelled by a conformal dimension $\Delta$ and by the spin $s$ representation of the Lorentz group $SO(d,\mathbb{C})$. In four dimensions, we denote operators in the $(s,\bar s)$ representation by
\begin{equation}
    O_{\alpha_1 \dots \alpha_s,\dot\alpha_1 \dots \dot\alpha_{\bar s}}(x)\ .
\end{equation}
The relation between the mass of a field in AdS$_{d+1}$ and the conformal dimension of the corresponding boundary operator for a scalar and for a totally symmetric spin-$s$ field is \cite{DHoker:2002nbb}
\begin{equation}\label{mass conformal dimension}
\begin{aligned}
    m^2L^2&\,= \Delta(\Delta-d)\ ,\qquad s=0\ ,\\
    m^2L^2&\,= (\Delta+s-2)(\Delta-s-d+2)\ ,\qquad s\geq 1\ ,
\end{aligned}
\end{equation}
where $L$ is the radius of AdS and $d$ is the boundary dimension. Analytically continuing $L\to i L$, this formula gives the masses of bulk fields in dS$_{d+1}$ \cite{Deser:2003gw}.

The main class of operators studied in this work are twist-two operators, where the twist is defined as
\begin{equation}
    \tau=\Delta-\frac{s+\bar s}{2}\ .
\end{equation}
These operators are special for several related reasons. For four-dimensional CFTs, unitarity of the $SO(4,2)$ representations imposes the bounds \cite{Minwalla:1997ka} 
\begin{equation}
    \begin{aligned}
        \Delta&\,\geq 1+\frac{s+\bar s}{2}\ ,
        \qquad s=0 \text{ or } \bar s=0\ ,\\
        \Delta&\,\geq 2+\frac{s+\bar s}{2}\ ,
        \qquad s>0 \text{ and } \bar s>0\ .
    \end{aligned}
\end{equation}
Thus twist-two operators with $s,\bar s>0$ saturate the unitarity bound and therefore correspond to shortened multiplets. These operators obey the conservation equation
\begin{equation}\label{conservation}
    \partial^{\beta\dot\beta}
    O_{\beta\alpha_2\dots\alpha_s,\dot\beta\dot\alpha_2\dots\dot\alpha_{\bar s}}
    =0\ .
\end{equation}
This condition is particularly familiar in the traceless-symmetric case $s=\bar s$, where the lowest-spin examples are the conserved current $J$, and the stress tensor $T$, corresponding respectively to $s=1,2$. By contrast, chiral operators of type $(s,0)$ or $(0,\bar s)$ do not shorten at twist two. Similarly, a scalar does not shorten at twist two. Nevertheless, the twist-two scalar is still distinguished holographically: it corresponds to $\Delta=d/2$, which is the value at which the different near-boundary falloffs of the bulk scalar degenerate. In AdS, $m^2L^2= -d^2/4$ saturates the Breitenlohner--Freedman bound, and in dS, $m^2L^2=d^2/4$ it is the boundary between light (principal series) and heavy fields (complimentary series). This should be contrasted with the scalar that appeared in the three-dimensional twistor construction, which trivialised twist-1 operators \cite{Baumann:2024ttn, CarrilloGonzalez:2025qjk, Bala:2025gmz}. There, the scalar was naturally interpreted as the operator dual to a conformally coupled bulk scalar in the alternate quantisation\footnote{For AdS, both quantisations are available in the Breitenlohner--Freedman window $    -\frac{d^2}{4}\leq m^2L^2\leq -\frac{d^2}{4}+1$ where both falloffs are normalisable. In the alternate quantisation, the roles of source and response are exchanged relative to the standard quantisation.}. For a conformally coupled scalar one has
\begin{equation}\label{Delta conformal}
    \Delta_\pm=\frac{d\pm 1}{2}\ .
\end{equation}
We will show that the ambitwistor framework can naturally capture the boundary propagators associated with conformally coupled scalars as well.

\subsection{CFT correlators in embedding space}
The embedding-space formalism has proved extremely useful for constructing conformal correlators, since conformal symmetry becomes manifest. The key observation is that the conformal group in $d$ dimensions is isomorphic to the Lorentz group in $d+2$ dimensions \cite{Weinberg:2010fx}. Consequently, conformal transformations, which act nonlinearly in physical spacetime, become linear in embedding space. 

The embedding may be realised explicitly by considering the projective null cone, $X^2=0$, in a six-dimensional flat space. In spinor notation this uses the two-to-one homomorphism for the complexified conformal group in four dimensions: $SL(4,\mathbb C) \rightarrow \mathrm{SO}(6,\mathbb C)$. One may then write \cite{Simmons-Duffin:2012juh,Binder:2020raz}
\begin{equation}\label{eq:XtoBispinor}
    \begin{aligned}
        X^{AB}
        &=
        \frac14
        \Lambda^A_\alpha
        \Lambda^{\alpha B}\ , \quad 
        X_{AB}
        &=
        \frac14
        \tilde\Lambda_A^{\dot\alpha}
        \tilde\Lambda_{\dot\alpha B}=\frac{1}{2}\epsilon_{ABCD} X^{CD} ,
    \end{aligned}
\end{equation}
where $A,B=1,\ldots,4$ are $SL(4,\mathbb C)$ indices, $\alpha,\dot\alpha=1,2$ correspond to the two $SL(2,\mathbb C)$ factors of little group, and $\epsilon_{ABCD}$ is the fully antisymmetric invariant tensor of $SL(4,\mathbb C)$. The bispinors satisfy
\begin{equation}
\label{eq:bispinor_rel}
    \Lambda^A_{\alpha}
    \tilde\Lambda_{A}^{\dot\alpha} = 0 \ ,
\end{equation}
and our spinor conventions are summarised in Appendix~\ref{app:conventions}.

Spin can be efficiently encoded by introducing an auxiliary null vector $U_i^M$ satisfying  $U_i^2=U_i\cdot X_i=0$. Alternatively, in terms of bispinors, one introduces 
\be
U^{AB}=\Upsilon^{[A}\Upsilon^{* B]}={\frac{1}{2} }\epsilon^{ABCD}\tilde{\Upsilon}_{[C}\tilde{\Upsilon}^{*}_{D]} \ , 
\ee
which can be made manifestly null and transverse by choosing
\begin{align}
    \Upsilon^A&=\xi^\alpha \Lambda^A_\alpha \ , \quad X_{AB}\Upsilon^{*B}=\tilde{\Upsilon}_A  \ ,\\
    \tilde{\Upsilon}_A&=\sigma_{\dot{\alpha}}\tilde{\Lambda}_A^{\adot} \ , \quad X^{AB}{\tilde{\Upsilon}^{*}}_B={\Upsilon}^A \ , 
\end{align}
which implies that $\xi_\alpha={\tilde{\Upsilon}^*}_A\Lambda^A_\alpha \ , \ \sigma^{\adot}={\Upsilon}^{*A}\tilde{\Lambda}^{\adot}_A$. Contracting the operator with the polarization spinors
$\Upsilon^{A_1}\cdots\Upsilon^{A_s}\tilde{\Upsilon}_{B_1}\cdots\tilde{\Upsilon}_{B_{\bar s}}$
packages the $(s,\bar s)$ representation into a scalar generating function. Hence, an operator of conformal dimension $\Delta_i$ obeys the homogeneity condition
\begin{equation} \label{eq:scalings}
O_i(r_i\Lambda_i,r_i\tilde\Lambda_i,q_i\Upsilon_i,\bar q_i\tilde\Upsilon_i)=r_i^{-2\Delta_i} \ q_i^{s} \ \bar q_i^{\bar s} \ O_i(\Lambda_i,\tilde\Lambda_i,\Upsilon_i,\tilde\Upsilon_i) \ .
\end{equation}

For three-point functions, every parity-even conformally invariant structure at separated points may be built from the basic invariants \cite{Costa:2011mg}
\begin{equation}\label{invariantseven}
    \begin{aligned}
        X_{ij}
        &=
        -2 X_i\cdot X_j,
        \\
        H_{ij}
        &=
        -2\Big[
        ( U_i\cdot U_j)(X_i\cdot X_j)
        -
        ( U_i\cdot X_j)( U_j\cdot X_i)
        \Big],
        \\
        V_i
        &=
        V_{i,jk}
        =
        \frac{
        ( U_i\cdot X_j)(X_k\cdot X_i)
        -
        ( U_i\cdot X_k)(X_j\cdot X_i)
        }{
        X_j\cdot X_k
        }.
    \end{aligned}
\end{equation}

The embedding-space formalism applies in arbitrary dimensions, but parity-odd three-point structures are special to $d\leq4$. At three points, only the six vectors $X_i$ and $U_i$ are available, so contractions with the $(d+2)$-dimensional Levi-Civita tensor can be non-vanishing only if $d+2\leq6$. Thus, four dimensions is the highest dimension admitting such structures, and in $d=4$ the corresponding parity-odd invariant is unique and given by
\begin{equation}\label{parity odd}
    \epsilon(X_1,X_2,X_3,U_1,U_2,U_3)\propto I_{12}I_{23}I_{31} + I_{13}I_{21}I_{32}.
\end{equation}
where 
\begin{equation}
 I_{ij} = \tilde{\Upsilon}_{iA}\Upsilon_j^A
\end{equation}
which is neither parity even nor parity odd by itself, and satisfies $H_{ij}=\frac12 I_{ij}I_{ji}$.  We note that in four dimensions, the basic three-point building blocks obey dimension-specific algebraic identities, so the resulting tensor structures are not all independent; we refer to \cite{Elkhidir:2014woa} for the explicit relations.

The embedding-space formalism therefore makes conformal symmetry manifest from the outset. However, the homogeneity condition, Eq.~\eqref{eq:scalings} with $\Delta=2+(s+\bar{s})/2$, is a necessary but not sufficient condition to construct a correlator of conserved currents. After constructing all conformally invariant tensor structures, one must additionally impose conservation. In embedding space, this takes the form
\begin{equation}\label{conservation embedding}
    \begin{aligned}
        (\partial_{X_i}\cdot D_i)
        \langle O_1\cdots O_n\rangle
        &=
        0,
        \\
        D_{iA}
        &=
        \left(
        \frac d2-1
        +
         U_i\cdot
        \frac{\partial}{\partial U_i}
        \right)
        \frac{\partial}{\partial U_i^A}
        -
        \frac12
         U_{iA}
        \frac{\partial^2}{
        \partial U_i\cdot\partial U_i
        }.
    \end{aligned}
\end{equation}

In practice, this approach requires constructing a large space of conformally invariant structures before imposing conservation as an additional constraint. Moreover, although solvable in principle, the conservation equations \eqref{conservation embedding} rapidly become cumbersome for higher-point functions and operators of large spin. It is therefore natural to seek a formalism in which conservation is implemented automatically from the outset. It has been shown that in the case of three-dimensional CFTs, twistor space can provide such framework \cite{Baumann:2024ttn,CarrilloGonzalez:2025qjk,Bala:2025gmz,Bala:2025jbh,Bala:2025qxr,Ansari:2025fvi}. As we will see below, ambitwistor space will be the appropriate generalization for the four-dimensional case.

\subsection{Ambitwistor space} \label{sec:ambitwistors}

In this section, we briefly review ambitwistor space, emphasising the aspects most relevant for our purposes. For further details, see \cite{Baston:1987av,Mason:2005kn,Mason:2013sva, Adamo:2016rtr,Seet:2024vmh,Geyer:2014fka}.

Ambitwistor space may be understood as the space of complex null geodesics in complexified Minkowski spacetime. In four dimensions, it is defined as the following quadric inside $\mathbb{CP}^3 \times \mathbb{CP}^3$:
\begin{equation}
    \mathbb{A}_{d=4}
    =
    \left\{
    (Z^A_i,W_{i,A}) \in \mathbb{CP}^3 \times \mathbb{CP}^3
    \;\middle|\;
    Z_i \cdot W_i =0
    \right\}.
\end{equation}
Here $Z^A$ and $W_A$ are respectively called twistors and dual twistors, transforming in the fundamental and anti-fundamental representations of $SL(4,\mathbb C)$, the complexification of the conformal group in 4d. We have included an index $i$ on the twistor and dual twistors for later convenience, since they will correspond to insertions of local operators $O_i$. As often when working with twistors, it will prove easier to work on complexified spacetime and pick a reality condition at the end. 
It is useful to introduce the homogeneous coordinates
\begin{equation}
    Z_i^A=(\lambda_i^\alpha,\mu_{i,\dot\alpha}),
    \qquad
    W_{i,A}=(\omega_{i,\alpha},\pi_i^{\dot\alpha}),
\end{equation}
with $\lambda^\alpha, \mu_{\dot \alpha}, \omega_\alpha, \pi^{\dot \alpha}$ two-component complex Weyl spinors.  The spinors $\lambda_i^\alpha$ and $\pi_{i,\dot\alpha}$ are homogeneous coordinates on two copies of $\mathbb{CP}^1$ transforming in the fundamental representations of the two $SL(2,\mathbb C)$ factors of the complexified Lorentz group. 

To make contact with spacetime, we impose the incidence relations
\begin{equation} \label{eq:incidence}
    \mu_i^{\dot\alpha}
    =
    x_i^{\alpha\dot\alpha}\lambda_{i,\alpha},
    \qquad
    \omega_i^\alpha
    =
    -x_i^{\alpha\dot\alpha}\pi_{i,\dot\alpha},
\end{equation}
where $x_i^{\alpha\dot\alpha}$ are coordinates on the physical conformally flat manifold $M_{4,\mathbb C}$, corresponding later to the insertion points of local operators $O_i\equiv O(x_i)$. Once the incidence relations are imposed, the constraint $Z_i\cdot W_i=0$ follows automatically.

Since the fundamental and anti-fundamental representations of
$SL(4,\mathbb C)$ are inequivalent, there is no conformally invariant operation that raises or lowers twistor indices. Introducing an infinity twistor $I_{AB}$ provides such a map only at the cost of choosing additional structure: it selects a conformal scale and breaks the full conformal symmetry to the subgroup that preserves $I_{AB}$. For flat space, one has
\begin{equation}\label{flat infinity}
    Z_i I^{\mathrm{flat}} Z_j
    =
    \langle \lambda_i \lambda_j \rangle,
    \qquad
    W_i I^{\mathrm{flat}} W_j
    =
    [\pi_i \pi_j].
\end{equation}

A useful alternative description is obtained by writing ambitwistors satisfying the incidence relations using the  bispinors in Eq.~\eqref{eq:XtoBispinor} as  
\begin{equation}\label{bispinors}
Z_i^A=\Lambda^A_{i\alpha}\lambda_i^\alpha, \qquad W_{i,A}=\tilde\Lambda_{i,A\dot\alpha}\pi_i^{\dot\alpha} \ .
\end{equation}
The condition $Z_i\cdot W_i=0$ then becomes equivalent to Eq.~\eqref{eq:bispinor_rel}. 

\paragraph{Conformal generators in ambitwistor space}
Just as the conformal group acts linearly in embedding space, its action
also linearises in ambitwistor space. Starting from the natural $\mathfrak{gl}(4)$ action on a single
ambitwistor pair $(Z_i^A,W_{i, A})$, one defines
\begin{equation}
T^A_{i \ B}
=
Z_i^A\frac{\partial}{\partial Z_i^B}
-
W_{i, B}\frac{\partial}{\partial W_{i, A}}.
\end{equation}
Its trace is
\begin{equation}
T^A_{i\ A}
=
Z_i^A\frac{\partial}{\partial Z_i^A}
-
W_{i, A}\frac{\partial}{\partial W_{i, A}},
\end{equation}
which measures the relative homogeneity between $Z_i$ and $W_i$. Since the conformal algebra is $\mathfrak{su}(2,2)$, or equivalently
$\mathfrak{sl}(4,\mathbb C)$ after complexification, one should project out the
trace and work with the traceless generators
\begin{equation}
\widetilde T^A_{i \ B}
=
T^A_{i \ B}
-\frac14 \delta^A_B\,T^C_{i \ C} \ , \label{eq:conf_gen_twist}
\end{equation}
which furnish the conformal action. The quadratic Casimir is given by
\begin{equation}
\mathcal C_{i,  2}
=
\widetilde T^A_{i \ B}\widetilde T^B_{i \ A}=
T^A_{i \ B}T^B_{i \ A}
-\frac14 \bigl(T^A_{i \ A}\bigr)^2 \ .
\end{equation}

\section{Conserved spinning correlators in ambitwistor space}\label{sec:conserved-spinning-ambitwistor}
Ambitwistor space differs conceptually from ordinary twistor space \cite{Adamo:2017qyl}; rather than encoding on-shell fields, it naturally describes off-shell,
divergence-free fields \cite{Baston:1987av}. These divergence-free fields, or conserved currents, are obtained by what we will refer to loosely as the ambitwistor Penrose transform:
\begin{equation}\label{ambitwistorPT}
    \psi^{s,\bar s}(x)
    =
    \oint_{\mathbb{CP}^1\times \mathbb{CP}^1}
    D\lambda\,D\pi\;
    \langle \xi\lambda\rangle^s
    [\sigma\pi]^{\bar s}
    f(Z,W)\big|_X \ ,
\end{equation}
where $\big|_X$ denotes that the ambitwistors are taken with the incidence relation in Eq.~\eqref{eq:incidence}, and the natural $\mathbb{CP}^1 \times \mathbb{CP}^1$ measure $D\lambda D\pi \equiv \braket{\lambda d\lambda}\braket{\pi d\pi}$ carries weight $2$ in $\lambda$ and $\pi$. For convenience, all spinor indices have been contracted with arbitrary
non-vanishing polarisation spinors $\xi^\alpha$ and $\sigma^{\dot\alpha}$. When these spinors are not contracted unto the Penrose transform above, we will refer to it as unpolarised. For our purposes, $f(Z,W)$ can be thought of locally as an holomorphic function of ambitwistors satisfying
\begin{equation}\label{scaling ambitwistor}
    f(rZ,\tilde rW)
    =
    r^{-s-2}\tilde r^{-\bar s-2}
    f(Z,W) \ ,
\end{equation}
which ensures that the integral is well-defined projectively. More precisely, $f(Z,W)$ represents a \v{C}ech class in $H^2(\mathbb A,\mathcal O(-s-2,-\bar s-2))$. We review this below, emphasizing how the chosen \v{C}ech representative fixes the contour up to homology. Readers already familiar with the cohomological formulation of the Penrose transform may skip this summary.

Without introducing any additional structure, this immediately suggests the following ansatz for the $n$-point correlator of twist-two operators of arbitrary spin:
\begin{equation}\label{general correlator formula}
    \boxed{
    \begin{aligned}
        \Big\langle
        \prod_{i=1}^n
        O_i^{s_i,\bar s_i}
        \Big\rangle
        =
        \prod_{i=1}^n
        \oint
        D\lambda_i\,D\pi_i\;
        \langle \xi_i\lambda_i\rangle^{s_i}
        [\sigma_i\pi_i]^{\bar s_i}
        \,
        f(Z_j\cdot W_k) \big|_X \ .
    \end{aligned}}
\end{equation}
Here $f(Z_j\cdot W_k)$ is the twistor-space correlator, translated to position space by nested ambitwistor Penrose transforms. This representation trivialises the kinematics, namely, it makes conformal invariance, conservation, and parity manifest.

\subsection{Ambitwistor Penrose transform and cohomology classes} \label{sec:PenroseT}

It was shown in \cite{Baston:1987av} that the cohomology group
$H^2(\mathbb A,\mathcal O(-s-2,-s-2))$ is isomorphic to the space of off-shell divergence-free fields of spin $(s,s)$ for $s>0$, and to the space of scalar fields with $\Delta=2$ for $s=0$. Here, $H^2(\mathbb A,\mathcal O(-s-2,-\bar s-2))$ denotes the \v{C}ech cohomology of locally holomorphic functions on projective ambitwistor space with projective weights $(-s-2,-\bar s-2)$ in the two ambitwistor variables $(Z,W)$. In a \v{C}ech description, one chooses an open cover $\{U_i\}$ of $\mathbb A$, and a class is represented by holomorphic functions on multiple overlaps of this cover, modulo \v{C}ech coboundaries. Equivalently, the same classes may be represented in Dolbeault form by $\bar\partial$-closed forms modulo $\bar\partial$-exact ones \cite{Penrose:1986ca,Ward:1990vs}.

More explicitly, projective ambitwistor space may be viewed topologically as $\mathbb{CP}^1\times\mathbb{CP}^1\times\mathbb C^3$ \cite{Baston:1987av}. Since the non-compact $\mathbb C^3$ factor may be restricted to a contractible coordinate patch, the non-trivial patching data is carried by the $\mathbb{CP}^1\times\mathbb{CP}^1$ directions, which can be covered by three Stein open patches. Thus, a class in $H^2(\mathbb A,\mathcal O(-s-2,-\bar s-2))$ can be represented by holomorphic functions $f_{ijk}$ on triple overlaps $U_i\cap U_j\cap U_k$. Representatives related by lower-degree overlap data are identified, schematically $f_{ijk}\sim f_{ijk}+f_{ij}+f_{jk}-f_{ik}$, where the $f_{ij}$ are holomorphic functions on double overlaps. Thus, the cohomology class is not a single globally defined function, but local holomorphic data together with its patching information.

The ambitwistor Penrose transform in Eq.~\eqref{ambitwistorPT} is the pushforward in the usual double fibration \cite{Adamo:2017qyl}. One pulls the cohomology representative back to the correspondence space using the incidence relations~\eqref{eq:incidence}, and then integrates over the $\mathbb{CP}^1\times\mathbb{CP}^1$ fiber above each spacetime point. In the \v{C}ech picture, the pulled-back representative is a cocycle on overlaps of this fiber, and the contour is chosen inside the relevant overlaps, where the representative is holomorphic. Deforming the contour without crossing singularities does not change the result by Cauchy's theorem, while changing the representative by a \v{C}ech-exact term gives contour integrals of holomorphic functions on patches, which vanish. Hence the spacetime field depends only on the cohomology class. The pole or contour choices made below are therefore not additional arbitrary choices: they are part of the local \v{C}ech representative, specifying the relevant overlaps and the contour pairing used to evaluate the Penrose transform.

\subsection{Manifest kinematic properties and symmetries}

\label{sec:manifest-kinematics}

\paragraph{Conformal invariance}

When constructing ambitwistor space correlators, we want to consider conformally invariant building blocks. Such building blocks are given by the twistor-dual twistor contractions
\begin{equation}
t_{ij}:=Z_i\cdot W_j  \qquad i\neq j  \ , \label{eq:tij}
\end{equation}
where the restriction $i\neq j$ follows from the ambitwistor constraint
$Z_i\cdot W_i=0$. In terms of spinors, we have
\begin{equation}
    t_{ij}= \braket{\lambda_i \omega_j}+ [\pi_j \mu_i]\ .
\end{equation}
It is easy to see that these bilinears are manifestly invariant under the diagonal action of the conformal generators:
\begin{equation}
    \sum_{k=1}^n
    \tilde{T}^{A}_{k\ B}
    (Z_i\cdot W_j)
    =
    0 \ .
\end{equation}
Hence, we will construct ambitwistor space correlators of the form
\begin{equation}
    f(Z_j, W_k)=f(Z_j\cdot W_k) \ .
\end{equation}
Consequently, all conformal covariance is carried entirely by the
Penrose-transform measure and spinning factors, while the representative
$f(Z_i\cdot W_j)$ itself is manifestly conformally invariant.

\paragraph{Conservation}

Let us now show that conservation becomes manifest in the ambitwistor formalism. As reviewed above, a necessary condition for a spin-$(s,s)$ operator to be conserved is that its conformal dimension satisfies
\begin{equation}
    \Delta
    =
    2+\frac{s+ \bar s}{2} \ . \label{eq:conserved_delta}
\end{equation}
Using the decomposition in Eq.~\eqref{bispinors}, and keeping the fiber coordinates $\lambda,\pi$ fixed, the Penrose transform scales under $X\rightarrow rX$ as
\begin{equation}
    \psi^{s,\bar s}(rX)=r^{-2-\frac{s+\bar s}{2}}\psi^{s,\bar s}(X).
\end{equation}
Here the fiber coordinates are inert under embedding-space rescalings, so this scaling follows from the homogeneity condition \eqref{scaling ambitwistor}. Comparing with Eq.~\eqref{eq:scalings} gives $\Delta=2+s$, as expected for symmetric traceless conserved currents when $s=\bar s$.

Conservation itself also becomes immediate in ambitwistor space. Applying the chain rule to the spacetime divergence $\partial/\partial x^{\alpha\dot\alpha}$ and using the incidence
relations gives
\begin{equation}\label{d/dx}
    \frac{\partial}{\partial x^{\alpha\dot\alpha}}
    =
    \pi_{\dot\alpha}
    \frac{\partial}{\partial\omega^\alpha}
    -
    \lambda_\alpha
    \frac{\partial}{\partial\mu^{\dot\alpha}}
    +\mathrm{c.c.}\ ,
\end{equation}
where $\mathrm{c.c.}$ stands for complex conjugate. The ambitwistor representative is holomorphic, that is, it depends only on $(Z,W)$ and not on their complex conjugates, so the conjugate terms drop out. The remaining derivatives contract with the spinor factors in the unpolarised Penrose transform and give
$\langle\lambda\lambda\rangle = [\pi\pi] = 0$. Thus the divergence vanishes already at the integrand level, yielding Eq.~\eqref{conservation}. This shortening is special to the $s,\bar s \neq 0$ case. For a chiral field $\psi_{\alpha_1\cdots\alpha_s}$, one term in Eq.~\eqref{d/dx} still contributes, so the twist-two condition is not a shortening condition; similarly, the scalar has no analogous conservation condition, as discussed in Section~\ref{sec:twist2}.

\paragraph{Parity at three-points}

In four dimensions, parity exchanges the two chiral spinor representations. In spinor-helicity variables this corresponds to exchanging the left- and right-handed spinors. At the level of the representative in ambitwistor space, the analogous operation exchanges twistors and dual twistors,
\begin{equation} \label{eq:parity}
Z_i \leftrightarrow W_i .
\end{equation}
Therefore, given a three-point correlator written in ambitwistor variables, its parity transform is obtained by applying this exchange to each external leg. More precisely, since the Penrose transform is specified in practice by both an integrand and a contour prescription, the operation in \eqref{eq:parity} should be understood as acting on the corresponding cohomology class. The integrand is transformed by exchanging $Z_i$ and $W_i$, and the contour is mapped to the parity-transformed contour. In practice, this amounts to mapping the pole prescription: a contour encircling a given pole in the original Z-variables is replaced by the contour encircling the parity-related pole in the W-variables. At the level of the representative, we define the parity-even projection by
\begin{equation}\label{parity evenness}
\left.
\braket{O_1^{s_1} O_2^{s_2} O_3^{s_3}}
\right|_{\mathrm{even}}
=
\frac{1}{2}
\left(
\braket{O_1^{s_1} O_2^{s_2} O_3^{s_3}}
+
(-1)^{s_1+s_2+s_3}
\left.
\braket{O_1^{s_1} O_2^{s_2} O_3^{s_3}}
\right|_{Z_i\leftrightarrow W_i}
\right) \  .
\end{equation}
Similarly, the parity-odd projection is
\begin{equation}\label{parity oddness}
\left.
\braket{O_1^{s_1} O_2^{s_2} O_3^{s_3}}
\right|_{\mathrm{odd}}
=
\frac{1}{2}
\left(
\braket{O_1^{s_1} O_2^{s_2} O_3^{s_3}}
-
(-1)^{s_1+s_2+s_3}
\left.
\braket{O_1^{s_1} O_2^{s_2} O_3^{s_3}}
\right|_{Z_i\leftrightarrow W_i}
\right) \ .
\end{equation}
At present, we regard \eqref{parity evenness} and \eqref{parity oddness} as an empirical rule for the three-point cohomology representatives. This is similar in spirit to the way twistor-space formulae for scattering amplitudes behave under parity transformations \cite{Arkani-Hamed:2009hub}. The rule passes several non-trivial checks that we performed for $s_i\leq 3$.

This projection is consistent with physical expectations. For example, let's consider the case of three non-abelian spin-one currents. The ambitwistor correlators we consider correspond to colour-ordered correlators. For parity-even bulk interactions, the vertices are proportional to 
$f^{a_1 a_2 a_3}$. Switching from the  DDM basis \cite{DelDuca:1999rs} to the trace basis implies the reflection identity $A_n[1,\cdots, n-1,n]=(-1)^n A_n[n, n-1,\cdots,1]$. This, together with the full permutation symmetry for the bosonic correlator, requires the relation in Eq.~\eqref{parity evenness}, since exchanging $Z_i\leftrightarrow W_i$ is equivalent to exchanging the operator insertions in our representatives (Eqs.~\eqref{mixed symmetry integrand all s}-\eqref{mixed symmetry opposite integrand all s}). In contrast, for the parity-odd structure, which can be viewed as arising from a five-dimensional Chern-Simons interaction, the vertices are proportional to the completely symmetric invariant tensor $d^{a_1 a_2 a_3}$. The associated three-point colour-ordered correlator is therefore symmetric under the exchange of two external legs, that is $A_3[1,2,3]=A_3[2,1,3]$. Thus, the required relation is now Eq.~\eqref{parity oddness}.

\subsection{Alternative representation: helicity sectors and derivative factors} 
The representation \eqref{general correlator formula} is not unique. In particular, there are two natural ways of realising the spin dependence at each insertion. One may either use explicit spinors, as in the multiplicative spinning factors above, or replace these factors by derivatives with respect to the conjugate twistor variables. We will refer to these two choices as the two \emph{helicity sectors} of a given insertion. This terminology is meant in the twistor sense: it labels the two natural chiral realisations of the spin degrees of freedom, rather than an on-shell four-dimensional helicity of the CFT operator.

Besides the multiplicative spinning factors used above, one may consider a purely derivative representative. For an operator of spin $(s_i,\bar s_i)$, this takes the form
\begin{equation}
    \Big\langle
        \prod_{i=1}^n
        O_i^{s_i,\bar s_i}
    \Big\rangle = \prod_{i=1}^n
    \oint
    D\lambda_i\,D\pi_i
    \left\langle \xi_i
        \frac{\partial}{\partial\omega_i}
    \right\rangle^{s_i}
    \left[\sigma_i \frac{\partial}{\partial\mu_i}\right]^{\bar s_i}
    F(Z_j\cdot W_k).
\end{equation}
Conservation is again manifest. Acting on the unpolarised Penrose transform with the divergence gives a factor
\begin{equation}
    \frac{\partial}{\partial\omega_i^\alpha}
    \frac{\partial}{\partial\mu_i^{\dot\alpha}}
    \left(
        \pi_i^{\dot\alpha}
        \frac{\partial}{\partial\omega_{i,\alpha}}
        -
        \lambda_i^\alpha
        \frac{\partial}{\partial\mu_i^{\dot\alpha}}
    \right)F .
\end{equation}
Both of these vanish because the derivatives commute. Thus the purely derivative representative is conserved directly at the integrand level, just like the purely multiplicative representative.

Mixed representatives, with both explicit spinors and derivatives at the same insertion, do not make conservation manifest. For example, a spin-$(2,2)$ insertion with one factor of each type is
\begin{equation}
    \Psi_{\alpha_1\alpha_2\dot\alpha_1\dot\alpha_2}
    =
    \oint
    D\lambda\,D\pi\;
    \lambda_{(\alpha_1}
    \frac{\partial}{\partial\omega^{\alpha_2)}}
    \pi_{(\dot\alpha_1}
    \frac{\partial}{\partial\mu^{\dot\alpha_2)}}
    F .
\end{equation}
Applying the divergence gives terms in which the contracted indices come from different types of factors. Schematically, one obtains non-vanishing contributions of the form 
\begin{equation}
    (\lambda_{\alpha_1}
\pi^{\dot\alpha_1}
\pi_{\dot\alpha_2}
\frac{\partial}{\partial \omega_{\alpha_1}}
\frac{\partial}{\partial \omega^{\alpha_2}}
\frac{\partial}{\partial \mu^{\dot\alpha_1}}
-
\lambda^{\alpha_1}
\lambda_{\alpha_2}
\pi_{\dot\alpha_1}
\frac{\partial}{\partial \omega^{\alpha_1}}
\frac{\partial}{\partial \mu_{\dot\alpha_1}}
\frac{\partial}{\partial \mu^{\dot\alpha_2}})
F .
\end{equation}
Such mixed representatives may still be conserved for special choices of $F$, and examples of this type do occur. However, conservation is then not automatic term by term. We therefore restrict to the two manifest sectors: the purely multiplicative and purely derivative representatives. For a generic $n$-point conserved correlator this gives $2^n$ helicity sectors, corresponding to the independent choice at each insertion of the multiplicative or derivative factor, while for identical external operators permutation symmetry reduces the inequivalent sectors to $n+1$.

\section{Two-and three-point functions}\label{sec:three-point-functions}
In this section, we explain how to obtain propagators and any conserved three-point correlator. We postpone the discussion of the two-point function to the end, since it involves an additional subtlety concerning conformal invariance, which we will explain there. We begin with the 3-point scalar case, since it already illustrates most of the essential features of the construction. The extension to spinning operators will then follow rather straightforwardly. As we will see, the main difficulty in this formalism lies in finding an admissible representative that does not integrate to zero. If the answer is non-zero and finite, by the discussion in \ref{sec:PenroseT}, it will automatically be in the space of conserved correlators. 

\subsection{Constructing the ambitwistor correlators: scalar case}

For scalar operators with $n=3$, Eq.~\eqref{general correlator formula} gives
\begin{equation}
    \braket{O_1O_2O_3}
    =
    \oint D\pi_{123}\, D\lambda_{123}\,
    f(Z_i,W_j)\big|_X \ ,
\end{equation}
where the ambitwistor correlator, or cohomology class representative, $f$ must have homogeneity $-2$ in each twistor $Z_i$ and dual twistor $W_j$. This homogeneity condition does not uniquely determine the form of the representative. For any function $F$, the function
\begin{equation}
    f=
    \frac{F\left(u\right) }{(Z_1\cdot W_2)(Z_2\cdot W_3)(Z_3\cdot W_1)(Z_1\cdot W_3)
    (Z_2\cdot W_1)(Z_3\cdot W_2)}
    \ , 
\end{equation}
where $u$ is the cross ratio given by
\begin{equation}
 u= \frac{
    (Z_1\cdot W_2)(Z_2\cdot W_3)(Z_3\cdot W_1)}{(Z_1\cdot W_3)(Z_3\cdot W_2)(Z_2\cdot W_1)} \ , \label{eq:cross_ration}
\end{equation}
has the required homogeneity properties. This shows that we have a cross-ratio already at three-points in ambitwistor space. For simplicity, let us consider $F(u)=u^q\ , $
with $q\in\mathbb{Z}^\ast$. It is easy to see that the corresponding correlator vanishes identically. Indeed, after performing the contour integrations, one necessarily encounters an integral around a second-order pole with no further dependence on the corresponding integration variable. The residue therefore vanishes.

The case $q=0$ is different. Although the representative has the correct naive projective homogeneity, it does not define a finite Penrose transform with this contour prescription. The divergence only appears at the last of the six contour integrations. Schematically, one obtains
\begin{equation}\label{divergent scalar}
    \frac{1}{X_{12}X_{23}X_{31}}
    \oint D\lambda\,\frac{1}{\braket{\lambda\lambda}} \ .
\end{equation}
Thus the ansatz produces the correct scalar three-point structure with $\Delta=2$ only up to an ill-defined overall factor, and should therefore be regarded as an inadmissible representative. Similarly, we have checked for several non-integer real values of $p$ that the resulting integral does not reproduce the correct answer, that is, it is not an admissible cohomology class representative.

This shows that the homogeneity requirement has to be understood with some care.  What matters for the correlation functions is not that the representative has the correct homogeneity pointwise as an integrand, but rather that it has the correct homogeneity inside the contour integral. In particular, the representative may transform by an additional term whose integral vanishes. This is precisely what logarithms allow. Under a rescaling, $ \log(rx)=\log x+\log r+2\pi i k$, $k \in \mathbb{Z}$, the anomalous term is independent of the integration variable. If the corresponding log-free integrand integrates to zero, then this anomalous contribution is harmless\footnote{Strictly speaking, the logarithms take us outside $H^2(\mathbb A,\mathcal O(-2,-2))$, whose representatives are holomorphic sections of fixed projective weight on overlaps. Their cohomological interpretation has been considered in \cite{Bailey:1985,Woodhouse_1988,Mason:1990FATT1}.}. These logarithms have been used previously to construct cohomology class representatives \cite{Bailey:1985,Woodhouse_1988,Mason:1990FATT1,Mason:2009sa,CarrilloGonzalez:2025qjk, Garner:2024tis}. 

There is, however, an important constraint. In the \v{C}ech description, the local representative must be holomorphic on the overlap where the contour is taken. Introducing a logarithm changes this local analytic structure, because one must choose a branch on the relevant $\mathbb{CP}^1$. Thus the contour must lie in a patch on which the logarithm is single-valued. In particular, the pole around which the contour is taken cannot lie on the branch cut of the logarithm. Equivalently, if the representative contains a factor $\log(Z_i\cdot W_j)$, then it should not also contain $(Z_i\cdot W_j)$ in the denominator. This immediately rules out modifying the $p=0$ representative above by simply inserting logarithms in the numerator.

The next simplest possibility is therefore to start from the $p=1$ representative. In this case, only three of the six possible factors $Z_i\cdot W_j$ appear as poles. We may then insert a logarithm of one of the remaining three factors. Under a rescaling, this logarithm shifts by a constant, and the resulting anomalous term is proportional to the log-free $p=1$ integrand, whose contour integral vanishes. The insertion is therefore consistent with the required projective homogeneity inside the integral.

One finds, however, that inserting a single logarithm still gives a vanishing result. The same is true after inserting two logarithms. This is natural, since there is no reason to privilege any one of the three remaining structures. The first non-vanishing representative is obtained by inserting logarithms for all three of them. We are therefore led to
\begin{equation}\label{scalar}
    \begin{aligned}
        \braket{O_1O_2O_3}
        &=
        \oint D\pi_{123}\,D\lambda_{123}\,
        \frac{
        \log(Z_1\cdot W_2)\,
        \log(Z_2\cdot W_3)\,
        \log(Z_3\cdot W_1)}
        {
        \bigl(
        (Z_1\cdot W_3)
        (Z_2\cdot W_1)
        ( Z_3\cdot W_2)
        \bigr)^2
        } \ .
    \end{aligned}
\end{equation}

Let us briefly clarify the logic of the evaluation of these correlators. We choose a representative by specifying which poles are enclosed by the contour integrals, together with an iterated residue prescription for the $\mathbb{CP}^1\times\mathbb{CP}^1$ fibre variables $\lambda_i$ and $\pi_i$, this is part of the Čech cohomology class data. The integrations should be performed one ambitwistor at a time, since conformal covariance is guaranteed only after the full $\mathbb{CP}^1\times\mathbb{CP}^1$ fibre integral associated with each ambitwistor has been performed. We show how this is performed in practice for the scalar in Appendix~\ref{app:perform_penrose}.

\subsection{Three-point spinning ambitwistor correlators}
Having understood the scalar case, the general spinning case is obtained by adjusting the powers of the invariants  $t_{ij}= Z_i \cdot W_j$,  so that the integrand has the required projective homogeneity, up to the logarithmic factors discussed above. One can formulate this in terms of weight-shifting operators.

In twistor space, twist preserving weight-shifting operators take on a very natural form. Because the correlator should overall be bosonic, it suffices to focus on the operator increasing the spin by one. This can be done either by increasing two operators 
\begin{equation}
    \begin{aligned}
        O_i O_j \xrightarrow{\times \frac{\braket{\xi_j \lambda_j} [\sigma_i \pi_i]u^p}{t_{ji}}} \bar \psi_i \psi_j \ ,
    \end{aligned}
\end{equation}
or by increasing the spin of one operator
\begin{equation}
    \begin{aligned}
        O_i \xrightarrow{\times \frac{\braket{\xi_i \lambda_i} t_{jk} [\sigma_i \pi_i] u^p}{t_{ji} t_{ik}}} J_i \ .
    \end{aligned}
\end{equation}
Hence, because of projectiveness, these operators are completely fixed (up to the cross ratio) and simply act by multiplication. By successive application of these operators, we can obtain the general spinning case. 

We will see that there are two natural options for the \v{C}ech cohomology class representatives, which we call the two branches. For bosonic equal spins, we find that one option is the branch with a representative given by 
\begin{equation}\label{mixed symmetry integrand all s}
 f_{3,p}
=
u^{-p}
\frac{
\log(t_{13})\log(t_{32})\log(t_{21})
}{
(t_{12}
t_{23}
t_{31})^{2+s}
}
\end{equation}
The opposite branch is obtained by exchanging $Z\leftrightarrow W$ together with $s\leftrightarrow\bar s$: 
\begin{equation} \label{mixed symmetry opposite integrand all s}
\widetilde{ f}_{3,p}
=
u^{p}
\frac{
\log(t_{12})\log(t_{23})\log(t_{31})
}{
(t_{13}
t_{32}
t_{21})^{2+s}
} 
\end{equation}
In both cases, we have $p\geq 0$. 

These correlators are not restricted to these simple case. Rather, they allow for general mixed-symmetry balanced representations labelled by $(s,\bar s)$, where the balanced sector is defined as the condition
\begin{equation}
    s_1+s_2+s_3=\bar s_1+\bar s_2+\bar s_3 \ .
\end{equation}
We then have more generally for the first branch
\begin{equation}\label{mixed symmetry integrand}
\boxed{
\begin{aligned}
 f_{3,p}
&\,=
u^{-p}
\frac{
t_{32}^{\bar s_1-s_3}
t_{13}^{s_2-\bar s_3}
\log(t_{13})\log(t_{32})\log(t_{21})
}{
t_{12}^{2+s_1+s_2-\bar s_3}
t_{23}^{2+s_2}
t_{31}^{2+\bar s_1}
} \\
p\geq p_{min}&\,=\text{ Max}(0,\bar s_3-s_2, s_3- \bar s_1)
\end{aligned}}
\end{equation}
while the opposite branch gives
\begin{equation} \label{mixed symmetry opposite integrand}
\boxed{
\begin{aligned}
\widetilde{ f}_{3,p}&\,
=
u^{p}
\frac{
t_{23}^{s_1-\bar s_3}
t_{31}^{\bar s_2-s_3}
\log(t_{12})\log(t_{23})\log(t_{31})
}{
t_{13}^{2+s_1}
t_{32}^{2+\bar s_2}
t_{21}^{2+\bar s_1+\bar s_2-s_3}
} \\
 p\geq p_{min}&\,= \text{Max}(0, s_3-\bar s_2, \bar s_3-  s_1)\ .
\end{aligned}}
\end{equation}
It would be interesting to understand whether this construction can be extended to the more general unbalanced case. We already see that the ambitwistor formalism encodes spin very efficiently: passing from scalar to spinning correlators amounts essentially to adjusting powers in the representative. This is in sharp contrast with position space, where spinning structures are considerably more involved. Before moving on to more interesting examples, we summarise some simple examples in Table~\ref{unique structure examples table}. One can efficiently relate our representatives to position space by evaluating the Penrose transform numerically as explained in Appendix \ref{app:numericalevaluation}. In addition to the scalar case and the examples listed in Table~\ref{unique structure examples table}, we have checked explicitly that our formalism with gives the (unique) structure for $\braket{O^{(3)}OO}$, all three possible structures for $\braket{JJJ}$ (including two parity even ones and one parity odd one) as well as the three structures of $\braket{JJT}$ and all five possible structures for $\braket{TTT}$. 
\begin{table}[t]
\centering
\renewcommand{\arraystretch}{1.6}
\begin{tabular}{c|c|c}
\hline
Correlator & Ambitwistor & Embedding space  \\
\hline
$\braket{\bar\psi_1\psi_2O_3}$
&
$\displaystyle
\,
\frac{
\log(t_{12})\log(t_{23})\log(t_{31})
}{
t_{13}^2t_{21}^3t_{32}^2
}
u^{p}
$
&
$\displaystyle
\frac{I_{21}}{2(p+1)^2X_{12}^2X_{23}X_{31}}
$
\\
\hline
$\braket{O_1O_2J_3}$
&
$\displaystyle
\frac{
t_{12}\log(t_{12})\log(t_{23})\log(t_{31})
}{
t_{13}^3t_{21}^2t_{32}^3
}
u^{p-1}
$
&
$\displaystyle
(-1)^p\frac{  V_3}{p(p+1)X_{23}^2X_{31}^2}
$
\\
\hline
$\braket{J_1J_2O_3}$
&
$\frac{
t_{23}t_{31}\log(t_{12})\log(t_{23})\log(t_{31})
}{
t_{13}^3t_{21}^4t_{32}^3
}
u^p
$
&
$\displaystyle
(-1)^{p+1}\frac{H_{12}-2V_1V_2}{3(p+2)^2X_{12}^3X_{23}X_{31}}
$
\\
\hline
$\braket{T_1T_2O_3}$
&
$\frac{
(t_{23}t_{31})^2\log(t_{12})\log(t_{23})\log(t_{31})
}{
t_{13}^4t_{21}^6t_{32}^4
}u^p
$
&
$\displaystyle
(-1)^p 2\frac{H_{12}^2-8V_1V_2 H_{12}+4 V_1 V_2}{(p+3)^2X_{12}^5X_{23}X_{31}}
$
\\
\hline
\end{tabular}
\caption{Examples of simple three-point correlators in ambitwistor space and embedding space. As we consider higher spins, the embedding space expression become more involved, but the ambitwistor ones do not.}
\label{unique structure examples table}
\end{table}

\paragraph{Yang-Mills}
We now turn to the spin-one case, which is more interesting because it admits three independent tensor structures. In embedding-space notation, we may write
\begin{equation}
    \braket{J_1J_2J_3}
    =
    \frac{
    a_1(p)\,V_1V_2V_3
    +a_2(p)\bigl(V_1H_{23}+V_2H_{31}+V_3H_{12}\bigr)
    +a_3(p)\,I_{12}I_{23}I_{31}
    }
    {
    (X_{12}X_{23}X_{31})^2
    } \ .
\end{equation}
On the twistor side, varying the parameter $p$, which measures the powers of the cross-ratio, no longer merely changes the overall normalization. Instead, within a fixed branch sector, varying $p$ can move us through a non-trivial subspace of conformally allowed structures. For example, considering Eq.~\eqref{mixed symmetry integrand} and \eqref{mixed symmetry opposite integrand}, we have checked that the twistor correlators span the three independent conserved structures allowed in embedding space. However, the precise dependence of the coefficients $a_i(p)$ on $p$ is not transparent from the ambitwistor representative. Rather than displaying the result for three different values of $p$, it is more useful to organize the answer according to parity. The three independent spin-one structures decompose into two parity-even structures and one parity-odd structure. The parity-even sector can be further identified with the two standard bulk interactions: the Yang--Mills interaction and the higher-derivative $F^3$ interaction. The remaining parity-odd structure corresponds instead to a Chern--Simons interaction.
\begin{equation}
    \braket{J_1J_2J_3}
    =
    c_{\mathrm{YM}}\,\braket{J_1J_2J_3}_{\mathrm{YM}}
    +
    c_{F^3}\,\braket{J_1J_2J_3}_{F^3}
    +
    c_{\mathrm{CS}}\,\braket{J_1J_2J_3}_{\mathrm{CS}} \ .
\end{equation}
Let us start with the parity-odd structure. Since this structure is unique, the choice of $p$ can only affect its overall normalization. We may therefore choose, for simplicity, $p=0$. Then, using Eq.~\eqref{parity oddness}, we obtain the following ambitwistor-space representative:
\begin{equation}
    \braket{J_1J_2J_3}_{\mathrm{CS}}= \frac{ log(t_{13})  log(t_{21}) log(t_{32})}{(t_{12} t_{23} t_{31})^3}+ \frac{ log(t_{12})  log(t_{23}) log(t_{31})}{(t_{13} t_{21} t_{32})^3}\ .
\end{equation}
Now, from Eq.~\eqref{parity evenness}, the parity-even sector is represented by
\begin{equation}
    \braket{J_1J_2J_3}_{p,\mathrm{even}}
    =
    \frac{
    \log(t_{13})\log(t_{21})\log(t_{32})\,u^{-p}
    }
    {
    (t_{12}t_{23}t_{31})^3
    }
    -
    \frac{
    \log(t_{12})\log(t_{23})\log(t_{31})\,u^{p}
    }
    {
    (t_{13}t_{21}t_{32})^3
    } \ .
\end{equation}
Different choices of $p$ give different linear combinations of the two parity-even structures. However, the relation between $p$ and the physical basis, spanned by the Yang--Mills and $F^3$ structures, is not simple. We have checked that the correlators obtained for $p=0$ and $p=1$ are linearly independent. They therefore form a basis for the parity-even sector.

In this basis, the Yang--Mills structure, characterized by \cite{Freedman:1998tz,Lee:2023qqx}
\begin{equation}
    a_1=\frac{6}{5}a_2,
    \qquad
    a_3=0,
\end{equation}
is given by
\begin{equation}
    \braket{J_1J_2J_3}_{\mathrm{YM}}
    =
    \#\left[
    \braket{J_1J_2J_3}_{0,\mathrm{even}}
    -9\braket{J_1J_2J_3}_{1,\mathrm{even}}
    \right] \ .
\end{equation}
Similarly, the $F^3$ structure, characterized by
\begin{equation}
    a_1=6a_2,
    \qquad
    a_3=0,
\end{equation}
is given by
\begin{equation}
    \braket{J_1J_2J_3}_{F^3}
    =
    \#\left[
    \braket{J_1J_2J_3}_{0,\mathrm{even}}
    +3\braket{J_1J_2J_3}_{1,\mathrm{even}}
    \right] \ .
\end{equation}
As explained in the appendix, these combinations were obtained by matching to the standard embedding-space tensor structures. It would be more satisfactory to understand the physical meaning of the parameter $p$ directly, so that the Yang--Mills and $F^3$ combinations could be derived a priori. In particular, a better understanding of the role of the cross ratio would clarify whether these physical structures can be associated with distinguished correlators with a given power of the cross ratio (individual values of $p$), or whether they are intrinsically realized only as linear combinations of different representatives.

\paragraph{GR}
The spin-two case proceeds in a closely analogous way. In this case, there are five independent conserved structures in total: three parity-even and two parity-odd. We have checked explicitly that varying the powers of the cross ratio (the value of $p$) produces five independent correlators. Again, rather than displaying all of them, let us focus only on the structure corresponding to general relativity.

In embedding-space notation, this structure is given by \cite{Zhiboedov:2012bm,Diwakar:2021juk}
\begin{equation}
\begin{aligned}
\braket{T_1T_2T_3}_{\mathrm{GR}}
&=
\frac{1}{8(X_{12}X_{23}X_{31})^3}
\Big[
16\,H_{12}H_{23}H_{31}
-54\,V_1^2V_2^2V_3^2
\\
&\qquad
-33\,V_1V_2V_3
\bigl(
H_{23}V_1+H_{31}V_2+H_{12}V_3
\bigr)
\\
&\qquad
+3\bigl(
H_{13}H_{23}V_1V_2
+H_{12}H_{32}V_1V_3
+H_{21}H_{31}V_2V_3
\bigr)
\\
&\qquad
+23\bigl(
H_{23}V_1
+H_{31}V_2
+(H_{12}+2V_1V_2)V_3
\bigr)^2
\Big] .
\end{aligned}
\end{equation}
From Eq.~\eqref{parity evenness}, the corresponding parity-even ambitwistor representatives take the form
\begin{equation}
\begin{aligned}
\braket{T_1T_2T_3}_{p,\mathrm{even}}
&=
\frac{
\log(t_{13})\log(t_{21})\log(t_{32})\,u^{-p}
}{
(t_{12}t_{23}t_{31})^4
}
+
\frac{
\log(t_{12})\log(t_{23})\log(t_{31})\,u^{p}
}{
(t_{13}t_{21}t_{32})^4
} .
\end{aligned}
\end{equation}
The three representatives with $p=0,1,2$ span the full three-dimensional parity-even sector. Matching to the embedding-space expression above, we find
\begin{equation}
\begin{aligned}
\braket{T_1T_2T_3}_{\mathrm{GR}}
=
\#
\bigg[
\frac{27}{4}\,
\braket{T_1T_2T_3}_{0,\mathrm{even}}
+
\frac{189}{8}\,
\braket{T_1T_2T_3}_{1,\mathrm{even}}
+
\frac{3375}{32}\,
\braket{T_1T_2T_3}_{2,\mathrm{even}}
\bigg] .
\end{aligned}
\end{equation}

Unfortunately, this expression is not as simple as the natural ambitwistor representatives in Eq.~\eqref{mixed symmetry integrand} and \eqref{mixed symmetry opposite integrand}, and it is not obtained by a direct double copy of the $\braket{J_1J_2J_3}_{\mathrm{YM}}$ representative discussed above. We suspect that there should be a more canonical choice of representatives within the relevant cohomology classes, both for the Yang--Mills and gravitational interactions, for which the double copy structure becomes manifest. This would be analogous to what was found on the Grassmannian side in \cite{Bala:2026trw}. Let us emphasize, however, that unlike that construction, the ambitwistor formula above directly produces the full position-space correlator, rather than a Wightman function from which the full correlator must subsequently be reconstructed from its discontinuity.

\subsubsection{Two-twistor representation}
So far, we have used ambitwistors $(Z_i^A,W_{iA})$ to construct the correlators. This makes conformal symmetry manifest as reviewed in Sec.~\ref{sec:manifest-kinematics}. One can instead pass to a representation involving two twistors\footnote{These are simply the so-called massive twistors \cite{Perjes, Albonico:2022pmd}.}, $(Z_i^A,Y_i^A)$, by Fourier transforming the dual twistor,
\begin{equation}\label{W to Y}
\tilde f(Y)
=
\int d^4W e^{iY\cdot W} f(W) \ .
\end{equation}
This is analogous to the relation between the purely twistorial and ambidextrous formulations of scattering amplitudes \cite{smatrix,Mason:2009sa}. The advantage of this representation is that it makes the twistor-space support of the correlator more transparent, in direct analogy with the twistor-string analysis of scattering amplitudes \cite{Witten:2003nn}, at the cost of manifest conformal invariance. 

In the twistor-string description of Yang-Mills amplitudes, an N$^k$MHV amplitude at loop order $l$ is supported on curves of degree $d=k+1+l$. So for instance, tree-level NMHV amplitudes are supported on degree 2 curves. In the scalar three-point correlator studied here, we find a closely related geometric structure: 3-point correlators can be written as a sum of reducible degree 3 curves in twistor space. To make this precise, let us introduce the weighted collinearity distribution imposing $Z_1$ to be restricted to the projective line through $Z_2$ and $Z_3$:
\begin{equation}
\tilde \delta^{(2)}_{p,q}(Z_1;Z_2,Z_3)
=
\int
\frac{ds}{s^{1+p}}
\frac{dt}{t^{1+q}}
\delta^4(Z_1-sZ_2-tZ_3) \ ,
\end{equation}
where the superscript denotes the codimension of the support of this condition, while the subscripts encode the projective weights, explicitly $Z_1$ has weight $-4-p-q$ while $Z_2$ and $Z_3$ have weights $p$ and $q$ respectively. The Fourier transform and the distribution above are to be understood as real integrals, so we need to choose a reality condition in which the ambitwistors themselves are real. This is done by considering the $(2,2)$, split spacetime signature, in contrast with the complexified coordinates used throughout the rest of this paper. As discussed in \cite{CarrilloGonzalez:2025qjk}, the passage from the complexified expression to its real counterpart is straightforward. For the scalar case, this simply replaces logarithmic functions by sign functions, and simple poles by first derivatives of Dirac delta functions. The ambitwistor expression for the scalar correlator \eqref{scalar} therefore becomes
\begin{equation}
f_3(Z_i,W_i)
=
\mathrm{sgn}(Z_2\cdot W_1)
\mathrm{sgn}(Z_3\cdot W_2)
\mathrm{sgn}(Z_1\cdot W_3)
\delta'(Z_1\cdot W_2)
\delta'(Z_2\cdot W_3)
\delta'(Z_3\cdot W_1)\ .
\end{equation}
Then using the Schwinger parameterisation of the sgn and $\delta'$ functions:
\begin{equation}
sgn(x)\,= \frac{1}{2\pi } \int_{-\infty}^{\infty} \frac{dc}{c} e^{-icx}\ , \ \quad 
    \delta'(x)\,= \frac{i}{2 \pi}
    \int_{-\infty}^{\infty} dc c e^{-icx}\ ,
\end{equation}
we obtain that the Fourier transform \eqref{W to Y} gives
\begin{equation}
\begin{aligned}
\tilde f_3(Z_i,Y_i)
&=
\tilde \delta^{(2)}_{-2,0}(Y_1;Z_3,Z_2)
\tilde \delta^{(2)}_{-2,0}(Y_2;Z_1,Z_3)
\tilde \delta^{(2)}_{-2,0}(Y_3;Z_2,Z_1).
\end{aligned}
\end{equation}
whose support is given by Fig. \ref{fig:triangle}

\usetikzlibrary{calc}

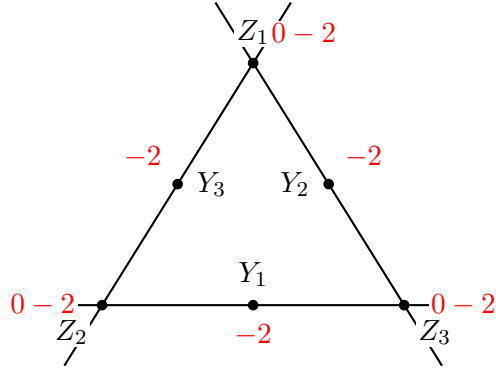
\begin{figure}[ht]
\centering
\begin{tikzpicture}
\coordinate (Z1) at (0,2);
\coordinate (Z2) at (-2,-1.2);
\coordinate (Z3) at (2,-1.2);

```
\coordinate (Y3) at (-1,0.4);   
\coordinate (Y2) at (1,0.4);    
\coordinate (Y1) at (0,-1.2);   

\draw[thick] ($(Z1)!-0.25!(Z2)$) -- ($(Z2)!-0.25!(Z1)$);
\draw[thick] ($(Z2)!-0.25!(Z3)$) -- ($(Z3)!-0.25!(Z2)$);
\draw[thick] ($(Z3)!-0.25!(Z1)$) -- ($(Z1)!-0.25!(Z3)$);

\fill (Z1) circle (2pt);
\node[above=6pt, fill=white, inner sep=1pt] at (Z1) {$Z_1$};
\node[above right=8pt, red, fill=white, inner sep=1pt] at (Z1) {$0-2$};

\fill (Z2) circle (2pt);
\node[below left=6pt, fill=white, inner sep=1pt] at (Z2) {$Z_2$};
\node[left=9pt, red, fill=white, inner sep=1pt] at (Z2) {$0-2$};

\fill (Z3) circle (2pt);
\node[below right=6pt, fill=white, inner sep=1pt] at (Z3) {$Z_3$};
\node[right=9pt, red, fill=white, inner sep=1pt] at (Z3) {$0-2$};

\fill (Y3) circle (2pt);
\node[right=6pt, fill=white, inner sep=1pt] at (Y3) {$Y_3$};
\node[above left=7pt, red, fill=white, inner sep=1pt] at (Y3) {$-2$};

\fill (Y2) circle (2pt);
\node[left=6pt, fill=white, inner sep=1pt] at (Y2) {$Y_2$};
\node[above right=7pt, red, fill=white, inner sep=1pt] at (Y2) {$-2$};

\fill (Y1) circle (2pt);
\node[above=6pt, fill=white, inner sep=1pt] at (Y1) {$Y_1$};
\node[below=6pt, red, fill=white, inner sep=1pt] at (Y1) {$-2$};
```

\end{tikzpicture}
\caption{Support of the scalar correlator in twistor space. The contributions to the projective weight of each twistor are shown in red.}\label{fig:triangle}
\end{figure}

This support is non-trivial. Six generic points in projective twistor space do not lie in a common projective plane, and a fortiori they do not lie on three lines in that plane. Here, however, the three collinearity constraints force the six twistors $Z_i,Y_i$ to lie on the three sides of the triangle determined by $Z_1,Z_2,Z_3$. This is reminiscent of the three-coplanar-line support found in the BCFW representation of NMHV Yang-Mills amplitudes in twistor space \cite{Mason:2009sa}, and also of the twistor-space support of one-loop NMHV box coefficients \cite{Bern:2004bt}. Finally, it is also natural to view this two-twistor representation as adapted to the five-dimensional massless little group $SO(3,\mathbb{C}) \simeq SL(2, \mathbb{C})/\mathbb{Z}_2$. The pair $(Z_i,Y_i)$ can therefore be packaged into a little-group doublet $Z_i^{A a}$.
In this language, the Fourier transform from $W_i$ to $Y_i$ trades the ambitwistor pair $(Z_i,W_i)$ for a pair of twistors carrying a manifest $SL(2,\mathbb{C})$ little-group index\footnote{Note that these do not correspond to the twistors obtained from a "naive" half-Fourier transform, used in \cite{Bala:2026trw} since these have factors $\delta(Z_i\cdot W_i)$, which are forbidden by the definition of ambitwistor space $Z_i\cdot W_i=0$ identically.}.

\subsection{Two-point spinning ambitwistor correlators}\label{sec:two-point-functions}
The two-point case differs from the three-point case in an important way. At three points, one can form a non-trivial projective cross-ratio out of the invariants $Z_i\cdot W_j$. At two points, no such conformally invariant cross-ratio exists. The only conformally invariant contractions are
\begin{equation}
    Z_1\cdot W_2,
    \qquad
    Z_2\cdot W_1 .
\end{equation}

Let us first construct the scalar two-point representative by following the same logic as in the scalar three-point case. The only manifestly conformally invariant representative with homogeneity $-2$ in each twistor and dual twistor is
\begin{equation}
    f_{2,0}
    =
    \frac{1}{
    (Z_1\cdot W_2)^2
    (Z_2\cdot W_1)^2
    } \ .
\end{equation}
However, this representative integrates to zero. Indeed, the relevant integration variable appears only through a double pole, so the corresponding residue vanishes. Moreover, because the pole cannot lie on the branch cut of a logarithm, one cannot cure this by simply dressing the same representative with factors such as $\log(Z_i\cdot W_j)$.

Without introducing additional data on ambitwistor space, the natural remaining possibility is to use the spinor contractions
\begin{equation}
    \braket{\lambda_1\lambda_2}=Z_1IZ_2 \  ,
    \qquad
    [\pi_1\pi_2]=W_1IW_2 \ ,
\end{equation}
which can be written in terms of ambittwistors by using the infinity twistor. This leads to the representative
\begin{equation}
\label{2 points}
\begin{aligned}
    \braket{O_1O_2}
    &=
    \oint D\pi_{12}\,D\lambda_{12}\,
    \frac{
    \log(Z_1IZ_2)\,
    \log(W_1IW_2)
    }{
    (Z_1\cdot W_2)^2
    (Z_2\cdot W_1)^2
    }
    \\
    &=
    \#\,\frac{1}{X_{12}^2} \ .
\end{aligned}
\end{equation}
The evaluation proceeds by the same contour-integral steps described for the three-point function and detailed in Appendix~\ref{app:perform_penrose}.

It is striking that the twistor representative depends explicitly on the infinity twistor, even though the final answer is conformally covariant. This is closely analogous to the situation for scattering amplitudes \cite{smatrix,Mason:2009sa}. In that context, the four-point Yang--Mills amplitude can be written entirely in terms of conformal invariants of the form $Z_i\cdot W_j$, while the three-point amplitude necessarily involves the infinity twistor. From a practical point of view, the reason is simple: with two twistors and one dual twistor, there are not enough conformally invariant contractions to build all required structures without using the infinity twistor. The correlator case is similar. At three points, we avoid this issue because we have in total 6 invariants $Z_i \cdot W_j$, but at two points we only have two and the infinity twistor appears to be unavoidable. The most general two-points can be constructed simply by introducing the spinning factors and adjusting the power for homogeneity
\begin{equation}\label{2 points mixed symmetry}
\boxed{
    \begin{aligned}
        \braket{O_1^{s_a, s_b} O_2^{s_b,s_a}}&\,= \oint D \pi_{12} D\lambda_{12} \braket{\xi_1 \lambda_1}^{s_a} [\sigma_1 \pi_1]^{s_b} \braket{\xi_2 \lambda_2}^{s_b}  [\sigma_2 \pi_2]^{s_a}  \frac{log(Z_1 I Z_2) log(W_1 I W_2)}{(Z_1 \cdot W_2)^{2+s_a}(Z_2 \cdot W_1)^{2+s_b}} \\
        &\,= \# \frac{I_{12}^{s_a}I_{21}^{s_b}}{X_{12}^{\Delta+\frac{s_a+s_b}{2}}} \ , 
    \end{aligned}}
\end{equation}

\section{Properties of the ambitwistor correlators} \label{sec:properties}
We now collect a few general properties of the ambitwistor correlators constructed above. In particular, the twistor representatives make the double copy structure of three-point functions and the form of the Ward identities especially transparent.
\subsection{Double copy}\label{sec:doublecopy}

The homogeneity of the ambitwistor Penrose transform leads to a simple double copy structure for three-point correlators. The statement is most transparent at the level of twistor representatives\footnote{The issues of interpreting the double copy for more than just representatives have been highlighted in the twistor space, classical double copy literature \cite{White:2020sfn, Beetar:2024ptv, CarrilloGonzalez:2022ggn, White:2024pve, Albertini:2025ogf}.}. Let $f_s$ denote the representative for the equal-spin three-point correlator. Then the spin-two representative may be written as
\begin{equation}
    f_2
    =
    \frac{f_1\tilde f_1}{\tilde f_0}\ .
\end{equation}
up to multiplication by a power of the cross-ratio $u^p$, that can always be chosen by picking the cross-ratio in $\tilde f_0$.  Here the scalar representative $\tilde f_0$ and $\tilde f_1$ are in the same branch. Meanwhile $f_1$ and $f_2$ are also in the same branch, which does not have to be that of $\tilde f_s$. With this, the ratio above carries precisely the projective weights required for the spin-two representative.

For example, choosing $\tilde f_0$ to be in the Eq.~\eqref{mixed symmetry integrand} branch. Then, the $\tilde f_s$ is
\begin{equation}
    \tilde f_s
    =
    \frac{
    \log(t_{13})\log(t_{32})\log(t_{21})
    }{
    (t_{12}t_{23}t_{31})^{s+2}
    }\ .
\end{equation}
The second single copy representative may then be chosen either on the same branch, $ f_s=\tilde  f_s$, or on the opposite branch,
\begin{equation}
    f_s
    =
    \frac{
    \log(t_{12})\log(t_{23})\log(t_{31})
    }{
    (t_{13}t_{32}t_{21})^{s+2}
    }\ .
\end{equation}
Thus, in this basis, any equal-spin spin-two three-point correlator can be obtained as a double copy of spin-one data. However, this double copy should not be identified with the (A)dS bulk double copy as the GR representative is not obtained from Yang--Mills one. 

Since the argument only uses the projective homogeneity of the representatives, the same construction extends to balanced mixed-symmetry three-point correlators. Namely, one may write
\begin{equation}
    f_{s_{2,i},\bar s_{2,i}}
    =
    \frac{
    f_{s_{1,i},\bar s_{1,i}}\,
    \tilde f_{\tilde s_{1,i},\bar{\tilde s}_{1,i}}
    }{
    \tilde f_0
    }\ ,
\end{equation}
where the spins add at each insertion,
\begin{equation}
    s_{2,i}
    =
    s_{1,i}
    +
    \tilde s_{1,i},
    \qquad
    \bar s_{2,i}
    =
    \bar s_{1,i}
    +
    \bar{\tilde s}_{1,i}.
\end{equation}
Again, the branch of one single-copy factor must be chosen compatibly with the branch of the scalar denominator. The analogous statement also holds at two points, but it is essentially trivial, as the two-point structure is unique and fixed completely by conformal symmetry. Indeed, the same factorisation already holds in the general embedding-space expression without having to rely on any special ambitwistor simplification.

\subsection{Ward identity}\label{sec:ward-identity}
In this section, we verify explicitly as a useful consistency check, that the $\braket{OOJ}$ correlator satisfies the expected non-homogeneous Ward identity
\cite{Baumann:2020dch}
\begin{equation}\label{Ward deltas}
    \begin{aligned}
          \nabla^{x_3}_\mu \braket{O_1 O_2 J^\mu_3}\propto \delta^4(x_3 -x_1) \braket{(\delta O_1) O_2}+\delta^4(x_3 -x_2) \braket{O_1 (\delta O_2)}\ .
    \end{aligned}
\end{equation}
In order to make these delta functions manifest, we simply integrate the lhs of Eq. \eqref{Ward deltas} over $x_3$ and use the divergence theorem to write it as a sphere integral centered on either $x_1$ or $x_2$, which is the same method that was used in 3d \cite{CarrilloGonzalez:2025qjk}. An $(n-1)$-sphere of radius $r$ embedded in $\mathbb R^n$ has surface element
\begin{equation}
    d\Sigma^\mu_{S^{n-1}}
    =
    n^\mu r^{n-1}d\Omega_{n-1},
    \qquad
    \int d\Omega_{n-1}
    =
    \frac{2\pi^{n/2}}{\Gamma(n/2)} ,
\end{equation}
where $n^\mu$ is the outward-pointing unit normal vector. In the present case, we consider a small three-sphere in the four-dimensional space parametrized by $x_3^\mu$, centered at $x_2$ and of radius $\epsilon$. Thus
\begin{equation}
    n^\mu=\frac{x_{23}^\mu}{\epsilon},
    \qquad
    d\Sigma^\mu_{S^3}
    =
    \epsilon^2 x_{23}^\mu d\Omega_3 \ .
\end{equation}
The sphere is chosen small enough to enclose $x_2$ but not $x_1$, and we take $\epsilon\to0$ at the end. By the divergence theorem, we obtain
\begin{equation}
\begin{aligned}
    \int d^4x\,\nabla_\mu \braket{O_1O_2J_3^\mu}
    &=
    \#
    \int d\Sigma^{\alpha\dot\alpha}_{S^3}
    \oint D\pi_{123}\,D\lambda_{123}\,
    \lambda_{3\alpha}\pi_{3\dot\alpha}\,
    \frac{
    t_{12}\log(t_{12})\log(t_{23})\log(t_{31})
    }{
    t_{13}^3t_{21}^2t_{32}^3
    }
    \\
    &=
    \#
    \epsilon^2
    \int d\Omega_3
    \oint D\pi_{123}\,D\lambda_{123}\,
    \langle \lambda_3 x_{23}\pi_3]\,
    \frac{
    t_{12}\log(t_{12})\log(t_{23})\log(t_{31})
    }{
    t_{13}^3t_{21}^2t_{32}^3
    } .
\end{aligned}
\end{equation}
We then perform the contour integrations in the same way as in the scalar case, which gives
\begin{equation}
\begin{aligned}
    \int d^4x\,\nabla_\mu \braket{O_1O_2J_3^\mu}
    &=
    \#
    \frac{\epsilon^2}{X_{12}}
    \int \frac{d\Omega_3}{X_{23}}\,
    \frac{x_{12}\cdot x_{31}}{X_{31}^2}
    \\
    &=
    \#
    \frac{\epsilon^2}{X_{12}}
    \int \frac{d\Omega_3}{X_{23}}\,
    \frac{(x_{13}-\epsilon n)\cdot x_{31}}{X_{31}^2} .
\end{aligned}
\end{equation}
Taking the limit $\epsilon\to0$, the last factor tends to $1/X_{31}$. Moreover,
\begin{equation}
    X_{23}=\epsilon^2 n^2=\epsilon^2 ,
\end{equation}
since $n^\mu$ is a unit vector. Therefore,
\begin{equation}
\begin{aligned}
    \int d^4x\,\nabla_\mu \braket{O_1O_2J_3^\mu}
    &=
    \#\,\frac{1}{X_{12}^2}
    \\
    &=
    \#\,\braket{O_1O_2} ,
\end{aligned}
\end{equation}
as expected from the non-homogeneous Ward identity. In three dimensions, one finds representatives both with and without logarithmic factors. It was observed that the latter representatives obey the homogeneous Ward identity, in the sense that the right-hand side of Eq.~\eqref{Ward deltas} vanishes even at coincident points. This provided a diagnostic for distinguishing different bulk interactions. In the present case, however, all our representatives contain logarithms, so this criterion no longer distinguishes between them, at least not in any direct way based on the presence of logarithmic factors.

Nevertheless, the fact that the Ward identity can be checked explicitly remains non-trivial. This is particularly clear when compared with the closely related Grassmannian formalism, which in three dimensions can be viewed as a Schwinger parametrisation of products of twistors. The Grassmannian representation, naturally trivialises the homogeneous Ward identity, rather than the full Ward identity including contact terms, although recent work has shown that, this is avoided in the presence of $\mathcal N=2$ supersymmetry \cite{Huang:2026tsh}.

\section{Beyond twist-two operators} \label{sec:beyond} 
The construction above was tailored to twist-two operators, for which the ambitwistor Penrose transform makes conservation manifest. We now ask how much of this structure survives away from the conserved point, where the corresponding bulk fields are massive and the correlators are no longer fixed by shortening.

\subsection{Towards massive fields}\label{sec:massive-fields}
As reviewed above, twistor methods are most naturally adapted to massless fields. Nevertheless, several extensions to massive fields have been explored; see for instance \cite{Hughston:1981,Eastwood:1981}. In particular, \cite{Hughston:1981} relates solutions of four-dimensional massive field equations to cohomology classes on a product of two twistor spaces $H^2(\mathbb{CP}^{3+} \times \mathbb{CP}^{3+}, \mathcal{O}_{m,s}(n,m))$. This is suggestive from our perspective, since the two-point representatives constructed in the previous sections already live naturally on two copies of ambitwistor space $H^2\big(\mathbb{PA}\times\mathbb{PA},\mathcal{O}(-s-2,-\bar s-2)\big)$. In this section, we show that the two-point functions associated with massive fields in the five-dimensional bulk can be obtained from a simple deformation of the two-point in Eq. \eqref{2 points mixed symmetry}.
At a practical level, the construction follows the same logic as in 3d \cite{CarrilloGonzalez:2025qjk, Bala:2025qxr}. Let us begin with the scalar case. For the twist-2 case, the ambitwistor representative is required to obey the homogeneity condition of Eq. \eqref{homogeneity ambitwistors}. Here, we simply generalise this to an arbitrary larger integer twist\footnote{Because of this condition, this means we are considering AdS massive fields rather dS ones, which do not take these values of the conformal dimension. We will show however in the next section that is possible to consider non-integer twist, at least for the conformally coupled scalar, which is relevant for both AdS and dS.} such that we now require 
\begin{equation}\label{homogeneity ambitwistors}
    f(rZ,rW)
    =
    r^{-2\Delta}
    f(Z,W) \ .
\end{equation}
Projective invariance of the Penrose transform then fixes the additional weight that must be inserted in the representative. Relative to the $\Delta=2$ case, this requires a factor of weight $\Delta-2$ in each of the $\lambda_1,\lambda_2,\pi_1,\pi_2$ variables. The simplest choice is
\begin{equation}
\big(\langle\lambda_1\lambda_2\rangle[\pi_1\pi_2]\big)^{\Delta-2}.
\end{equation}
This insertion uses the infinity twistor, see Eq.~\eqref{flat infinity}, and therefore breaks manifest conformal invariance. For integer $\Delta$, we are led to the representative
\begin{equation}\label{eq:massive-scalar-2pt}
    \begin{aligned}
        \braket{O_{1,\Delta}O_{2,\Delta}}
        &=
        \oint D\pi_{12}\,D\lambda_{12}\,
        \big((Z_1IZ_2)(W_1IW_2)\big)^{\Delta-2}
        \frac{\log(Z_1IZ_2)\log(W_1IW_2)}
        {(Z_1\cdot W_2)^{\Delta}(Z_2\cdot W_1)^{\Delta}}
        \\
        &=\#\,\frac{1}{X_{12}^{\Delta}} .
    \end{aligned}
\end{equation}
Thus the deformation away from the conserved value simply reproduces the expected scalar two-point function with arbitrary conformal dimension.

The spinning generalisation proceeds exactly as in the three-dimensional case \cite{CarrilloGonzalez:2025qjk}: this ansatz only reproduces the completely symmetric part of the correlator, and so we have to set $\xi_2=\xi_1$ and $\sigma_1=\sigma_2$. However, by conformal invariance, this component contains enough information to determine the full two-point function (in embedding space they look the same, so it is trivial to reconstruct the full propagator). For an operator carrying $(s_a,s_b)$ spinor indices, we find
\begin{equation}\label{eq:massive-spinning-2pt}
   \boxed{
   \begin{aligned}
        \braket{O_{1,\Delta}^{s_a,s_b}O_{2,\Delta}^{s_b,s_a}}_{\mathrm{sym}}
        &=
        \oint D\pi_{12}\,D\lambda_{12}\,
        \braket{\xi_1\lambda_1}^{s_a}
        [\sigma_2\pi_1]^{s_b}
        \braket{\xi_1\lambda_2}^{s_b}
        [\sigma_2\pi_2]^{s_a}
        \\
        &\qquad\times
        \big((Z_1IZ_2)(W_1IW_2)\big)^{\tau+2}
        \frac{\log(Z_1IZ_2)\log(W_1IW_2)}
        {(Z_1\cdot W_2)^{\Delta+(s_a-s_b)/2}
        (Z_2\cdot W_1)^{\Delta+(s_b-s_a)/2}}
        \\
        &=\#\,
        \frac{I_{12}^{s_a}I_{21}^{s_b}}
        {X_{12}^{\Delta+(s_a+s_b)/2}} ,
    \end{aligned}}
\end{equation}
where
\begin{equation}
    \tau=\Delta-\frac{s_a+s_b}{2}
\end{equation}
is the twist. When either $s_a=0$ or $s_b=0$, the symmetric component coincides with the full correlator.\\

Finally, there is another way to understand the result. Since the divergence operator admits a simple ambitwistor-space representation, as follows from Eq.~\eqref{d/dx}, the same is true of the Laplacian. Its action at the first operator insertion is given in terms of ambitwistor coordinates as
\begin{equation}
    \left[\pi_1\frac{\partial}{\partial\mu_1}\right]
    \left\langle\lambda_1\frac{\partial}{\partial\omega_1}\right\rangle  \ .
\end{equation}
This gives a direct ambitwistor-space interpretation of the familiar position-space identity
\begin{equation}
    \Box_1\braket{O_1(x_1)O_2(x_2)}_{\Delta}
    =
    4\Delta(\Delta-1)
    \braket{O_1(x_1)O_2(x_2)}_{\Delta+1}.
\end{equation}
Thus higher-twist two-point functions can be generated recursively from lower-twist ones by acting with the ambitwistor representative of the Laplacian. This construction reproduces the deformed representatives in Eq.~\eqref{eq:massive-spinning-2pt}. For instance applying the Laplacian on the twist-2 scalar we obtain
\begin{equation}
    \begin{aligned}
        &\left[\pi_1\frac{\partial}{\partial\mu_1}\right]
        \left\langle\lambda_1\frac{\partial}{\partial\omega_1}\right\rangle
        \frac{\log(Z_1IZ_2)\log(W_1IW_2)}
        {(Z_1\cdot W_2)^{2}(Z_2\cdot W_1)^{2}}
        \\
        &\qquad =
        (Z_1IZ_2)(W_1IW_2)
        \frac{\log(Z_1IZ_2)\log(W_1IW_2)}
        {(Z_1\cdot W_2)^{3}(Z_2\cdot W_1)^{3}}
        +\cdots \, .
    \end{aligned}
\end{equation}
where $\cdots$ stand for terms that contain less than two logarithms and integrate to 0. The RHS is then indeed the $\Delta=3$ scalar propagator, as expected from Eq. \eqref{eq:massive-spinning-2pt}. \\
From this perspective, the simplification at two points is tied to the equality $\Delta_1=\Delta_2$. The Laplacian acting on either insertion raises the common conformal dimension in a controlled way, so the result remains within the one-dimensional space of scalar two-point structures. At higher points this argument no longer works in the same way. Acting with $\Box_i$ does not simply raise $\Delta_i$ while leaving the other conformal dimensions fixed. Rather, because the result must be re-expanded in a basis of conformal three-point structures, the operation can mix the effective dimensions assigned to the other insertions as well. It therefore raises the total conformal weight in a controlled way, but does not generate the full space of correlators at the new twist.

More concretely, at $n$ points one can act independently with the $n$ Laplacians $\Box_i$. Starting from a given seed correlator, this produces at most $n$ independent higher-twist primaries. This is generally fewer than the number of independent conformal structures available at the same total twist. For example, starting from the scalar three-point function with $\Delta_i=2$, a single Laplacian produces scalar three-point functions with total dimension
\begin{equation}
    \Delta_1+\Delta_2+\Delta_3=8.
\end{equation}
Acting at each of the three insertions gives three structures. However, the integer-dimension scalar correlators with this total dimension include the permutations of
\begin{equation}
    (\Delta_1,\Delta_2,\Delta_3)=(2,3,3),
    \qquad
    (\Delta_1,\Delta_2,\Delta_3)=(2,2,4).
\end{equation}
These give six possibilities in total. The Laplacian construction therefore captures only half of the available space.

\subsection{Conformally coupled scalar}\label{sec:conformally-coupled-scalar}
In the previous section we explained how massive AdS bulk fields can be incorporated in our formalism. There, however, we restricted attention to integer conformal dimensions, since the Penrose transform could then be evaluated by ordinary residues. In this section we take a first step beyond this restriction.  The construction below does not directly follow the ordinary \v{C}ech-cohomology framework of Sec.~\ref{sec:PenroseT}, but should instead be viewed as a generalization of it. Although we do not prove an isomorphism between the relevant twisted cohomology classes and operators dual to conformally coupled scalar fields, the structure of the construction suggests that such an interpretation could exist.

Considering non-integer spins is useful not only as a technical extension of the formalism, but also for physical applications, since a conformally coupled scalar in $(A)dS_5$ has conformal dimensions (cf. Eq.~\eqref{Delta conformal})
\begin{equation}
\Delta_{+}=\frac{5}{2} \, \quad \Delta_{-}=\frac{3}{2} \  .
\end{equation}
It is therefore natural to ask whether the same ambitwistor-space construction can reproduce its propagator. We focus here on the $\Delta_+$ branch. In this case Eq. \eqref{eq:massive-scalar-2pt} gives
\begin{equation}\label{conformally coupled scalar}
\mathcal I_{5/2}
=
\oint D\pi_{12}\,D\lambda_{12}\,
\frac{\big(\braket{\lambda_1 \lambda_2}[\pi_1 \pi_2]\big)^{1/2}
\log(\braket{\lambda_1 \lambda_2})\log([\pi_1 \pi_2])}
{(t_{12}t_{21})^{5/2}}\ .
\end{equation}
To evaluate the integral, we introduce affine coordinates
\begin{equation}
\lambda_{i\alpha}=(1,z_i),
\qquad
\pi_{i\dot\alpha}=(1,w_i),
\end{equation}
and choose a frame in which
\begin{equation}
x_{12}^\mu=(r,0,0,0)\, ,
\end{equation}
where $r \in \mathbb{C}$. In these coordinates, the integral takes the form
\begin{equation}
\mathcal I_{5/2}
=
r^{-5}
\oint dw_2\,dz_2\,
J_{w_1}(w_2,z_2)\,J_{z_1}(z_2,w_2),
\end{equation}
where
\begin{equation}\label{eq:Jz-integral}
J_z(a,b)
=
\oint_{P_z} dz\,
\frac{(a-z)^{1/2}\log(a-z)}
{(1+bz)^{5/2}}\, .
\end{equation}
We see that the spacetime dependence factorises from the fibre and can be pulled outside the contour integrals. Thus, the remaining task is to show that the fibre integrals are finite and non-vanishing.

We choose $P_z$ to be a Pochhammer contour. As reviewed in Appendix~\ref{app:Pochhammer}, this is the natural replacement of an ordinary residue contour when the integrand is multivalued. For the integral above, the Pochhammer contour gives
\begin{equation}
J_z(a,b)
=
\frac{8\pi}{3}\,
\frac{b^{-3/2}}{1+ab}\ ,
\end{equation}
which is proven in the same appendix. Substituting this expression back into Eq.~\eqref{conformally coupled scalar}, we obtain
\begin{equation}
\mathcal I_{5/2}
=
\left(\frac{8\pi}{3}\right)^2
r^{-5}
\oint dw_2\,dz_2\,
\frac{(z_2w_2)^{-3/2}}
{(1+z_2w_2)^2}.
\end{equation}
The remaining integral is now an ordinary residue computation, which yields
\begin{equation}
\mathcal I_{5/2}
=
\frac{32\pi^2}{3}\,
\frac{1}{r^5}\propto
\frac{1}{(x_{12}^2)^{5/2}} \ ,
\end{equation}
as expected for a scalar two-point function of conformal dimension $\Delta=5/2$.

\section{Conclusions}
In this work we have developed an ambitwistor-space approach to four-dimensional CFT correlators. The central observation is that the ambitwistor Penrose transform naturally produces off-shell divergence-free fields \cite{Baston:1987av}, and is therefore particularly well suited to correlators of twist-two operators in complexified spacetime. These are relevant because they give both the late time wavefunction coefficients of photons and gravitons in dS$_5$ and their boundary correlator in AdS$_5$, by choosing appropriate reality conditions. Conformal invariance is built into the dependence on the basic twistor--dual-twistor contractions $Z_i\cdot W_j$, while conservation follows directly from the structure of the transform. Moreover, the Penrose transform separates the spin dependence from the remaining dynamical data (bulk interaction or equivalently, OPE coefficients), allowing correlators of different spin to be treated in a uniform way.

We constructed explicit \v{C}ech representatives for scalar and spinning two- and three-point functions, mirroring the previous construction of massless dS$_4$ correlators in twistor space. However, already in the scalar three-point case, a new feature of the four-dimensional construction appears: unlike in the three-dimensional twistor construction, ambitwistor space admits a non-trivial cross-ratio at three points. Consequently, the required projective weights do not uniquely fix the representative. In practice, the admissible representatives involve logarithms whose anomalous projective transformations integrate to zero, and therefore still define well-behaved projective Penrose transforms. Evaluating these representatives by iterated contour integrals reproduces the expected position-space correlators.

For spinning operators, the ambitwistor formulation gives a compact and uniform description of the allowed conserved structures. The dependence on spin is implemented by shifting the powers of the basic invariants and inserting the appropriate spinning factors. We showed in particular that the ambitwistor representatives span the allowed three-point structures for spin one and spin two.  We also conjectured a parity projection operation for the 3-points correlator directly in ambitwistor space. This operation should be viewed as acting on the cohomology class rather than just the representative, that is in practice on both the integrand and the contour of integration. Our formula was checked in several non-trivial cases but a better understanding of it would be desirable. The appearance of the parity-odd structures -- which we computed explicitly for the Chern-Simons correlator of three spin-1 currents -- is especially suggestive, since these include contributions usually associated with loop effects on the boundary \cite{Schreier:1971um,Coriano:2023hts}. Understanding how such information is encoded in ambitwistor space would be an interesting direction for future work. Furthermore, we also showed that one can transform our three-point functions from ambitwistor space to a two-twistor representation where the correlator localizes on degree three curves. This was checked explicitly in the scalar case.

The two-point functions require a slightly different ingredient. In contrast with the three-point case, there is no non-trivial conformal cross-ratio built purely from the invariants $Z_i\cdot W_j$, and the representatives naturally involve the infinity twistor. This mildly obscures conformal invariance at the level of the representative, even though the final answer is conformally covariant. This is reminiscent of the role played by the infinity twistor in flat-space twistor amplitudes, where certain low-point amplitudes cannot be written using conformal invariants alone. It would be useful to understand this breaking more intrinsically, and in particular whether it reflects an unavoidable feature of two-point data or merely a limitation of the representatives used here.

We also showed that the two-point construction can be deformed away from the conserved value: by inserting suitable powers of infinity-twistor contractions, one obtains scalar and spinning two-point functions with more general integer conformal dimensions, corresponding in AdS to massive bulk fields at the level of boundary propagators, though not to unitary dS fields. The same deformation can be understood through the ambitwistor representation of the Laplacian which raises the conformal dimension. This mechanism is special to two points, where conformal symmetry fixes the answer uniquely; at higher points, the Laplacian does not generate the full space of higher-twist structures, though more general deformations of the representatives may still exist. Finally, for the conformally coupled scalar in AdS/dS, ordinary contours are replaced by a Pochhammer contour closed in twisted homology, reproducing the expected two-point function with $\Delta=5/2$ and suggesting a natural extension of the ambitwistor Penrose transform to twisted cohomology, potentially relevant for more general massive fields.

As a consistency check, we verified the non-homogeneous Ward identity for the $\langle JOO\rangle$ correlator directly in ambitwistor space. The computation relates the divergence of the three-point representative to the scalar two-point representative, as expected. This confirms that the logarithmic representatives reproduce the full correlators, rather than only their homogeneous or discontinuity parts. This should be contrasted with the Grassmannian approach, which in the absence of supersymmetry naturally produces the discontinuity of the correlator. At the same time, the distinction between homogeneous and non-homogeneous solutions is less transparent here than in the three-dimensional construction, precisely because the representatives used in this work all involve logarithms. In addition, we showed that the twistor formalism naturally leads to a basis in which a boundary double copy structure is manifest.

One of the main lessons of this paper is that the natural cohomology class representatives for the ambitwistor correlators in Eqs.~\eqref{mixed symmetry integrand} and \eqref{mixed symmetry opposite integrand} do not correspond directly to individual bulk interactions. Rather, they provide a natural basis from which the bulk-interaction basis can be reconstructed. This is perhaps not surprising, since our construction was formulated intrinsically from the boundary, unlike the three-dimensional twistor construction of \cite{CarrilloGonzalez:2025qjk}, where the boundary twistor correlators were motivated by a boundary limit of bulk twistor data. In the present four-dimensional case, the boundary ambitwistors are closely related to the mini-twistors of the five-dimensional bulk. Nevertheless, the ambitwistor Penrose transform used here is distinct from the AdS$_5$ Penrose transform of \cite{Adamo:2016rtr}. It would be interesting to understand whether a bulk perspective following \cite{Adamo:2016rtr,Seet:2024vmh,Geyer:2020iwz} can lead to a more canonical choice of representatives, one more directly adapted to Yang--Mills, $F^3$, Einstein gravity, and higher-derivative bulk interactions.

A related open problem is to clarify the relation between the ambitwistor representatives constructed here and the Grassmannian formulation. Such a relation may shed light on the role of the three-point cross-ratio, suggest an analogue of little-group covariance in the ambitwistor description, and provide a more direct route to the non-homogeneous Ward identities in the Grassmannian approach. It may also help identify representatives in which the double copy structure is more closely aligned with the bulk double copy. We are currently exploring these directions.

A natural next step is to extend the construction to higher points, where one can test whether the simplifications found here continue to organise the much larger space of spinning conformal structures. Finally, since some boundary correlators in 4 dimensions have also been obtained from twistors and ambitwistor strings, it would be interesting to understand whether these two formulations are related \cite{Adamo:2017zkm, Eberhardt:2020ewh, Roehrig:2020kck, Caron-Huot:2023wdh}.

\section*{Acknowledgements}
We thank Mattia Arundine, Daniele Pavarini, Guilherme Pimentel, and Facundo Rost for many useful discussions and collaborations in related topics. We are also grateful to Lionel Mason for insightful discussions.  MCG is supported by the Imperial College Research Fellowship. TK is supported by an STFC studentship.

\appendix
\section{Conventions}\label{app:conventions}
Spinor indices are raised and lowered with
\begin{equation}
    \epsilon^{\alpha\beta}
    = -\epsilon_{\alpha\beta}=
    \begin{pmatrix}
    0 & 1\\
    -1 & 0
    \end{pmatrix},
\end{equation}
and similarly for dotted indices, with the brakets always defined in the NW-SE convention
\begin{equation}
    \braket{ab}= a^\alpha b_\alpha\ , [ab]=a^{\dot \alpha} b_{\dot \alpha}\ .
\end{equation}
We take the 4d sigma matrices to be
\begin{equation}
    (\sigma^0)^{\alpha \dot \alpha}
    =
    i
    \begin{pmatrix}
    1 & 0\\
    0 & 1
    \end{pmatrix},
    \qquad
    (\sigma^1)^{\alpha \dot \alpha}
    =
    \begin{pmatrix}
    1 & 0\\
    0 & -1
    \end{pmatrix},
\end{equation}
\begin{equation}
    (\sigma^2)^{\alpha \dot \alpha}
    =
    \begin{pmatrix}
    0 & 1\\
    1 & 0
    \end{pmatrix},
    \qquad
    (\sigma^3)^{\alpha \dot \alpha}
    =
    \begin{pmatrix}
    0 & -i\\
    i & 0
    \end{pmatrix}.
\end{equation}
The conjugate matrices are chosen as
\begin{equation}
    (\widetilde\sigma^0)_{\dot \alpha \alpha }=(\sigma^0)^{\alpha \dot \alpha},
    \qquad
    (\widetilde\sigma^i)_{\dot \alpha \alpha }=-(\sigma^i)^{\alpha \dot \alpha},
    \qquad
    i=1,2,3.
\end{equation}
With these conventions,
\begin{equation}
    \sigma^\mu \widetilde\sigma^\nu
    +
    \sigma^\nu \widetilde\sigma^\mu
    =
    -2\eta^{\mu\nu}\mathbf 1_2\ ,
\end{equation}
where $\eta^{\mu \nu}$ is the Euclidean metric in the plus-signature in these conventions.
A vector $x^\mu$ is converted to a spinor by
\begin{equation}
    x^{\alpha\dot\alpha}
    =
    -x_\mu(\sigma^\mu)^{\alpha\dot\alpha}.
\end{equation}
It satisfies
\begin{equation}
    \det(x^{\alpha\dot\alpha})
    =
    -x^2,
    \qquad
    x^\mu
    =
    \frac12
    \operatorname{tr}
    \left(
    x\,\widetilde\sigma^\mu
    \right), \qquad x_{\alpha\dot \alpha}
    =\epsilon_{\alpha\beta}  \epsilon_{\dot\alpha\dot\beta}  x^{\beta\dot\beta} .
\end{equation}
We embed the four-dimensional space into the six-dimensional projective null cone with metric
\begin{equation}
    \eta_{MN}
    =
    \operatorname{diag}(1,1,1,1,1,-1).
\end{equation}
The embedding coordinate is
\begin{equation}
    X^M
    =
    \left(
    x^\mu,
    \frac{1-x^2}{2},
    \frac{1+x^2}{2}
    \right),
\end{equation}
which obeys
\begin{equation}
    X^2=0.
\end{equation}
Similarly, a physical polarisation vector $z^\mu$ is lifted to
\begin{equation}
    U^M
    =
    \left(
    z^\mu,
    -z\cdot x,
    z\cdot x
    \right).
\end{equation}
This satisfies
\begin{equation}
    U\cdot X=0,
    \qquad
    U^2=z^2.
\end{equation}

We also use six-dimensional chiral sigma matrices $(\Sigma^M)^{AB}$ and
$(\widetilde\Sigma^M)_{AB}$, normalised by
\begin{equation}
    \Sigma^M\widetilde\Sigma^N
    +
    \Sigma^N\widetilde\Sigma^M
    =
    -2\eta^{MN}\mathbf 1_4.
\end{equation}
where the embedding space has signature $(+++++-)$.
In terms of the four-dimensional sigma matrices above, one convenient explicit realisation is
\begin{equation}
    \begin{aligned}
        \Sigma^0 &= -\sigma^2\otimes\sigma^3,
        &
        \Sigma^1 &= i\sigma^3\otimes\sigma^2,
        \\
        \Sigma^2 &= -i\sigma^3\otimes\sigma^1,
        &
        \Sigma^3 &= i\sigma^3\otimes\sigma^0,
        \\
        \Sigma^4 &= \sigma^0\otimes\sigma^3,
        &
        \Sigma^5 &= -i\sigma^1\otimes\sigma^3.
    \end{aligned}
\end{equation}
The conjugate matrices are
\begin{equation}
    \widetilde\Sigma^0=-\Sigma^0,
    \qquad
    \widetilde\Sigma^1=\Sigma^1,
    \qquad
    \widetilde\Sigma^2=\Sigma^2,
\end{equation}
\begin{equation}
    \widetilde\Sigma^3=-\Sigma^3,
    \qquad
    \widetilde\Sigma^4=\Sigma^4,
    \qquad
    \widetilde\Sigma^5=-\Sigma^5.
\end{equation}
A six-dimensional vector may be written as an antisymmetric bispinor by
\begin{equation}
    X_{AB}
    =
    \frac14
    X_M
    (\widetilde\Sigma^M)_{AB},
    \qquad
    X^{AB}
    =
    \frac14
    X_M
    (\Sigma^M)^{AB}.
\end{equation}
These matrices are antisymmetric,
\begin{equation}
    X_{AB}=-X_{BA},
    \qquad
    X^{AB}=-X^{BA},
\end{equation}
and the vector can be recovered from them through
\begin{equation}
    X^M
    = (\Sigma^M)^{AB}X_{AB}
    =(\widetilde\Sigma^M)_{AB}X^{AB}\ .
\end{equation}

\section{Performing the Penrose transform}
\label{app:perform_penrose}
In this appendix, we spell out how the contour prescription used in the main text is implemented both analytically and numerically.

\subsection{Contour prescription}
Here, we clarify the logic of the evaluation of the Penrose transform in detail for the scalar case.  In practice, we identify a  representative by keeping track of which poles are enclosed by the contour integrals, together with an iterated residue prescription for the $\mathbb{CP}^1\times\mathbb{CP}^1$ fiber variables $\lambda$ and $\pi$. This has a direct \v{C}ech interpretation: the local functions are defined on specified overlaps of charts, and the contours are taken inside those overlaps, where the representative is holomorphic. Here we use a product cover with two open Stein patches on each $\mathbb{CP}^1$, compatible with the K\"unneth decomposition\footnote{We should nevertheless remember that our representatives are not strictly in $H^2(\mathbb{CP}^1\times\mathbb{CP}^1,\mathcal O(a,b))$, thus a possible obstruction to this decomposition might exist. } \cite{Adamo:2017qyl}.
\begin{equation*}
	H^2(\mathbb{CP}^1\times\mathbb{CP}^1,\mathcal O(a,b))
	\simeq
	H^1(\mathbb{CP}^1,\mathcal O(a))
	\otimes
	H^1(\mathbb{CP}^1,\mathcal O(b)) \ .
\end{equation*}

As we now show explicitly, Eq.~\eqref{scalar} evaluates to the correct position-space correlator. An important practical point is that the integrations should be performed one ambitwistor at a time: for each insertion, the $\pi_i$ and $\lambda_i$ residues are taken successively before moving on to another insertion. Thus, we do not freely interchange the order of the nested ambitwistor Penrose transforms, for example by integrating $\pi_i$ and then $\pi_j$ before completing the full $\mathbb{CP}^1\times\mathbb{CP}^1$ fiber integral at insertion $i$. This is natural because conformal covariance is guaranteed only after the full fiber integral associated with each ambitwistor has been performed.

There is also a related caution in the \v{C}ech description. After a contour integral has been performed, the local form of the remaining representative can change, including the positions of the poles and branch cuts seen by the remaining variables. Thus intermediate expressions should not be interpreted as independent \v{C}ech representatives on the original cover. Rather, the nested calculation should be viewed as an iterated-residue realisation of the original Penrose transform, with the contour at each stage chosen in the overlap where the corresponding local representative is holomorphic.

Clearly, the initial choice of ambitwistor over which we integrate is arbitrary, so let us pick the third, which we can do as long as we pick a contour that does not cross the branch cuts of the logarithms:
\begin{equation}
	\begin{aligned}
		\braket{O_1O_2O_3}
		&=
		\oint D\pi_{123}\,D\lambda_{123}\,
		\frac{
			\log(Z_1\cdot W_2)\,
			\log(Z_2\cdot W_3)\,
			\log(Z_3\cdot W_1)}
		{
			\bigl(
			(Z_1\cdot \bm W_3)
			(Z_2\cdot W_1)
			(\bm Z_3\cdot W_2)
			\bigr)^2
		} \ .
	\end{aligned}
\end{equation}
Here, we denote in bold the poles around which we integrate for clarity. Performing the first ambitwistor Penrose transform gives
\begin{equation}\label{pole curly}
	\begin{aligned}
		\braket{O_1O_2O_3}_0&\,= \oint D\pi_{12} D\lambda_{12}  \frac{log(Z_1 \cdot W_2) }{ (Z_2 \cdot \bm{W}_1)^2} \frac{(\mathcal{Z}_3 \cdot W_1)}{(\mathcal Z_3 \cdot W_2) (W_1 \cdot X_3 \cdot W_2)} \frac{(Z_2 \cdot \mathcal W_3)}{(\bm{Z}_1 \cdot \mathcal W_3) (Z_2 \cdot X_3 \cdot Z_1)}\ ,
	\end{aligned}
\end{equation}
where by the curly variable we mean,
\begin{equation}
	\mathcal Z_i  \, \equiv Z_i|_{\lambda_i=l_{i}}\ , \,\, \mathcal{W}_i= W_i|_{\pi_i= m_{i}}\ ,
\end{equation}
where 
\begin{equation}\label{arbitrary spinors}
	l_{i}, m_{i}
\end{equation}
are arbitrary reference spinors, which appear after using the residue theorem. These arise simply because \begin{equation}\label{log Z nm}
	\begin{aligned}
		\oint D \lambda_i \frac{log(Z_i \cdot W_j)}{(Z_i \cdot W_k)^{2}}&\, = \frac{(\mathcal Z_i \cdot W_j)}{(\mathcal Z_i \cdot W_k)(W_k \cdot X_i \cdot W_j)}
	\end{aligned}
\end{equation} which follows from writing the double pole as the derivative of a simple pole
\begin{equation}
	\left(\frac{\braket{l_{i} \frac{\partial}{\partial\lambda_i}}}{\langle l_{i} x_{ik} \pi_k]}\right) \frac{1}{Z_i \cdot W_k}  \ ,
\end{equation}
and integrating by parts. Next, we can integrate around either ambitwistor again. If we proceed to integrate $Z_1, W_1$, the choice of pole for $W_1$ does not matter since the analytic structure is simply that of two poles and therefore it does not matter which we enclose.\footnote{This is true for the scalar. For spinning correlators, we are only guaranteed to obtain the correct answer when integrating around $(Z_2 \cdot \bm{W}_1)$. Thus, among the residue prescriptions considered here, this appears to be the admissible \v{C}ech realisation of the representative. It would be desirable to understand this restriction from first principles.} However the function on $\mathbb{CP}^1_{\lambda_1}$ has a more complicated analytic structure as before: it has a branch cut as well as two simple poles and therefore the choice of pole around which we integrate matters. It is imperative to integrate around $(Z_1 \cdot {\cal W}_3)$ rather than $(Z_2 \cdot X_3 \cdot Z_1)$, since the latter gives infinity. We obtain:
\begin{equation}
	\begin{aligned}
		\braket{O_1O_2O_3}_0&\,=  \oint D\pi_2 D \lambda_2 \frac{log( \mathcal W_3 \cdot X_1 \cdot W_2)}{(Z_2 \cdot X_1 \cdot X_3 \cdot \bm W_2)^2}\ .
	\end{aligned}
\end{equation}
At this stage, the remaining integral should be treated as an iterated residue, as mentioned earlier. In the coordinates and representative used here, the analytic structure in one $\mathbb{CP}^1$ variable changes after the other residue has been taken. Before the $\pi_2$ residue is evaluated, the dependence on $\lambda_2$ appears only through a double pole, so taking the $\lambda_2$ residue first gives zero. Performing the $\pi_2$ integral first instead turns the remaining $\lambda_2$ integral into one with simple poles, from which the desired residue is obtained. This does not mean that the ambitwistor Penrose transform itself is order-dependent. It means that the present product contour prescription and coordinate chart obscure the natural geometric contour. The singularities are supported on mixed divisors in $\mathbb{CP}^1\times\mathbb{CP}^1$, so the appropriate pairing should be understood using a contour adapted to those divisors, rather than as a product of one-dimensional residues. The computation above should therefore be viewed as an iterated-residue realisation of the appropriate \v{C}ech construction, rather than as an arbitrary choice of order. Nevertheless, it would be desirable to have a better first principles description to choose the representatives.

After performing the $\pi_2$ integral, the remaining integrand contains two simple poles in $\lambda_2$. We may then integrate around one of them, obtaining the position-space scalar correlator
\begin{equation}
	\begin{aligned}
		\braket{O_1O_2O_3}_0
		&=
		\frac{1}{X_{12}X_{31}}
		\oint D\lambda_2\,
		\frac{
			\mathcal W_3\cdot X_1\cdot \mathcal W_2}
		{
			(Z_2\cdot X_1\cdot X_3\cdot \mathcal W_2)
			(Z_2\cdot \mathcal W_3)
		} \\
		&=
		\frac{1}{X_{12}X_{23}X_{31}} \ .
	\end{aligned}
\end{equation}

Two final remarks are in order. First, since conformal symmetry already fixes the scalar three-point function uniquely, it is not necessary to impose parity evenness separately. In fact, one can check that the two terms in Eq.~\eqref{parity evenness} are equal in the scalar case.

Second, nothing fundamentally forces us to choose $p=0$ in Eq.~\eqref{mixed symmetry opposite integrand}. More generally, any positive integer value of $p$ is allowed, provided that the pole does not lie on the branch cut of the logarithm. Varying $p$ changes only the overall normalization of the answer, producing a factor proportional to $(-1)^p/(p+1)^2$ multiplying the correlator. A priori, one may introduce an arbitrary function of the cross-ratio. For a generic choice, however, this would introduce essential singularities, so we restrict to power-law dependence on the cross-ratio.

\subsection{Numerical method}
\label{app:numericalevaluation}
The ambitwistor Penrose transform involves six contour integrals. Rather than evaluating these integrals symbolically, we implement the contour prescription directly in Mathematica. We first introduce affine coordinates on each fibre,
\begin{equation}
    \lambda_i=(1,z_i),
    \qquad
    \pi_i=(1,w_i),
\end{equation}
and use the Euclidean conventions of Appendix~\ref{app:conventions}. 
With these conventions the Penrose transform is computed as an iterated contour integral. 
For each check we assign generic integer values to the external kinematic data: the entries of $(x_{ij})^{\alpha\dot\alpha}$, the polarisation spinors $\xi_i^\alpha$ and $\sigma_i^{\dot\alpha}$, and the auxiliary spinors appearing in Eq.~\eqref{arbitrary spinors}. The latter are parametrised, for example, as $l_{i}=(0,\tilde l_{i})$. The numerical values are generated in Mathematica inside a \texttt{Block}, using \texttt{RandomInteger} in the range $[-10^6,10^6]$. This produces generic kinematic points while avoiding accidental degeneracies. Since the Pauli matrices are complex in this basis, these checks should be understood as being performed at generic complex values of the spacetime data.

The six contour integrals are then evaluated by applying the pole prescription described in Section~\ref{sec:three-point-functions}. Concretely, at each step we solve for the prescribed pole in the relevant affine variable and extract the coefficient of the simple pole. For instance, a typical step has the form
\begin{equation}
\begin{aligned}
    &\text{\texttt{pole1 = z[6] /. Solve[t[1,3] == 0, z[6]] // DeleteDuplicates;}}\\
    &\text{\texttt{int1 = Residue[Frep, \{z[6], pole1[[1]], -1\}];}}
\end{aligned}
\end{equation}
where $\texttt{t[i,j]}$ denotes $t_{ij}$, as defined in Eq.~\eqref{eq:tij}, and $\texttt{Frep}$ is the integrand of the Penrose transform for a given spin configuration and power of the cross ratio $u$. Repeating this procedure for the six integration variables gives the value of the Penrose transform at the chosen kinematic point.

Finally, we compare this result with the most general embedding-space ansatz
\begin{equation}
    \left\langle \prod_i O_i \right\rangle
    =
    \sum_{i=1}^n a_i S_i\, ,
\end{equation}
evaluated on the same kinematic data. Here the $S_i$ are the independent allowed tensor structures. By repeating the computation for $n$ sets of random values, one obtains a linear system for the coefficients $a_i$. Solving this system fixes the expansion of the ambitwistor result in the embedding-space basis.
\section{Derivative sector at three-points}\label{app:derivative-sector}
For completeness, let us also state the result in the full derivative sector. We give, for example, the corresponding representative for traceless symmetric operators in the branch of Eq. \eqref{mixed symmetry opposite integrand}. It takes the form
\begin{equation}
    \begin{aligned}
        \braket{\tilde O^{s_1}_1\tilde O^{s_2}_2\tilde O^{s_3}_3}_p
        &=
        \#\oint D\pi_{123}D\lambda_{123}
        \prod_{i=1}^3
        \biggl(
        \left\langle\xi_i\frac{\partial}{\partial\omega_i}\right\rangle
        \left[\sigma_i\frac{\partial}{\partial\mu_i}\right]
        \biggr)^{s_i}
        \biggl[
        M_{2-s_1}(t_{13})
        M_{2-s_1-s_2+s_3}(t_{21})\\
        &\hspace{1.2cm}\times
        M_{2-s_2}(t_{32})
        t_{23}^{s_3-s_1}
        t_{31}^{s_3-s_2}
        \log(t_{12})\log(t_{23})\log(t_{31})
        u^p
        \biggr]  \ .
    \end{aligned}
\end{equation}
Here $p\geq p_{\min}=\mathrm{Max}(0,s_2-s_3,s_1-s_3)$, and we have defined
\begin{equation}
    M_n(f_{ij})=
    \begin{cases}
        f_{ij}^{-n}, & n>0,\\
        f_{ij}^{|n|}\log(f_{ij}), & n\leq 0.
    \end{cases}
\end{equation}

In three dimensions, the Yang--Mills correlator is most naturally written in a particular derivative sector. One might therefore hope that the simple four-dimensional expression found above has a similar origin but this is not the case. To illustrate this, we record in Table~\ref{derivatives JJJ} a few low $p$ examples for the parity-even spin-one correlator,
\begin{equation}
    \braket{J_1J_2J_3}_{\mathrm{even},p}
    =
    \frac{
    a_1 V_1V_2V_3
    +a_2\left(V_1H_{23}+V_2H_{31}+V_3H_{12}\right)
    }{
    \left(X_{12}X_{23}X_{31}\right)^2
    } ,
\end{equation}
obtained in different derivative sectors.
\begin{table}[H]
\centering
\renewcommand{\arraystretch}{1.2}
\begin{tabular}{c c c c}
\hline
$n$ & $k$ & $a_1$ & $a_2$ \\
\hline
$1$ & $k$ & $0$ & $(-1)^k/3$ \\
\hline
$2$ & $0$ & $-2$ & $-1$ \\
\hline
$2$ & $1$ & $2$ & $7/3$ \\
\hline
$2$ & $2$ & $-2$ & $-4$ \\
\hline
$3$ & $0$ & $-4$ & $-2$ \\
\hline
$3$ & $1$ & $20$ & $10$ \\
\hline
$3$ & $2$ & $-52$ & $-34$ \\
\hline
\end{tabular}
\caption{
Low $p$ derivative-sector representatives for the parity-even spin-one three-point function. The integer $n=\sum_i n_i$ counts the total number of derivative pairs appearing in the spinning factors, with $n_i$ defined by
$\left(\braket{\xi_i\frac{\partial}{\partial\omega_i}}[\sigma_i\frac{\partial}{\partial\mu_i}]\right)^{n_i}$.
The second column fixes the cross-ratio power through $p=p_{\min}+k$. The coefficients $a_1$ and $a_2$ determine the resulting linear combination of the two parity-even tensor structures. The Yang--Mills structure requires $a_1=\frac{6}{5}a_2$, whereas the $F^3$ structure requires $a_1=6a_2$. Thus, at low values of $p$, no individual derivative sector directly isolates either the Yang--Mills or the $F^3$ correlator.
}\label{derivatives JJJ}
\end{table}

\section{Dolbeault representatives}\label{app:dolbeault}
Conserved fields in spacetime can be represented by cohomology classes on ambitwistor space. As discussed in Section \ref{sec:conserved-spinning-ambitwistor}, there are two standard realisations of these classes: Čech and Dolbeault. In the main text, we have worked entirely in the Čech description. Although one expects an equivalent Dolbeault formulation of our results, the translation is not always transparent. In this appendix, we illustrate the idea in the simplest scalar example. The Dolbeault representative provides an alternative regularisation of Eq. \eqref{divergent scalar}.

The divergence in Eq. \eqref{divergent scalar} comes from the final integral. The basic idea is therefore to replace this contour integral by an integral over a compact space. In the Dolbeault description, this can be achieved by introducing the Euclidean conjugate twistor
\begin{equation}\label{Euclidean Conjugation twistor}
\begin{aligned}
    Z^A&\,=(\lambda_\alpha,\mu^{\dot{\alpha}})
    \xrightarrow{}
    \hat Z^A=(\hat \lambda_\alpha,\hat \mu^{\dot{\alpha}})\ ,
\end{aligned}
\end{equation}
where
\begin{equation}\label{euclidean conjugation}
\begin{aligned}
    \hat\lambda^\alpha&\,=(-\bar \lambda^1,\bar \lambda^0)\ ,\\
    \hat \mu^{\dot \alpha}&\,=(-\bar \mu^{\dot 1},\bar \mu^{\dot 0})\ .
\end{aligned}
\end{equation}
The holomorphic $\mathbb{CP}^1$ measure $D\lambda$ is then replaced by the $S^2$ measure $D\lambda \wedge D\hat \lambda$. Correspondingly, the representative is allowed to depend on both $Z_i\cdot W_j$ and $\hat Z_i\cdot W_j$, with the homogeneities adjusted so that the Penrose transform remains projectively well-defined.

For the scalar three-point function, this gives
\begin{equation}
\begin{aligned}
    \braket{O_1O_2O_3}
    &\,=
    \int_{S^2}
    \frac{D\lambda_1 \wedge D\hat \lambda_1}{\braket{\lambda_1 \hat \lambda_1}}
    \oint
    D\lambda_{23}D\pi_{123}
    \frac{1}{
    (Z_1 \cdot W_2)
    (\hat Z_1 \cdot W_3)
    (Z_2 \cdot W_3)
    (Z_2 \cdot W_1)
    (Z_3 \cdot W_1)
    (Z_3 \cdot W_2)
    }\ .
\end{aligned}
\end{equation}
After performing the contour integrals, one obtains
\begin{equation}
\begin{aligned}
    \braket{O_1O_2O_3}
    &\,=
    \frac{1}{X_{12}X_{23}X_{31}}
    \int_{S^2}
    \frac{D\lambda_1 \wedge D\hat \lambda_1}
    {\braket{\lambda_1 \hat \lambda_1}^2} \\
    &\,=
    \frac{4\pi}{X_{12}X_{23}X_{31}}\ .
\end{aligned}
\end{equation}
Thus the Dolbeault representative gives a finite answer without introducing the branch cuts that appeared in the Čech regularisation. The price is that the construction no longer looks holomorphic: it explicitly involves the conjugate twistor $\hat Z$. Moreover, once such variables are introduced, additional ratios such as
\begin{equation}
    \frac{Z_i\cdot W_j}{\hat Z_i\cdot W_j}
\end{equation}
can appear. This makes the structure of the representative less rigid and obscures a direct uniform generalisation to the full set of conserved correlators considered in the main text.

\section{The Pochhammer contour}\label{app:Pochhammer}
Let us first recall some basic facts about the Pochhammer contour. Consider
\begin{equation}
\oint_{P_u} u^{\alpha-1}(1-u)^{\beta-1}\,du .
\end{equation}
The integrand is multivalued, with branch points at $u=0$ and $u=1$, and generically also at $u=\infty$. A loop around $u=0$ changes the chosen branch by the monodromy factor $e^{2\pi i\alpha}$, while a loop around $u=1$ changes it by $e^{2\pi i\beta}$. Therefore a single loop around either branch point is not, in general, a closed contour on the local system defined by the integrand.

The Pochhammer contour is the commutator of the two elementary loops. If $\gamma_0$ is a positively oriented loop around $u=0$ and $\gamma_1$ is a positively oriented loop around $u=1$, both based at a point between $0$ and $1$, we take
\begin{equation}
P_u=\gamma_0\gamma_1\gamma_0^{-1}\gamma_1^{-1}\, .
\end{equation}
The total monodromy is then trivial:
\begin{equation}\label{monodromy}
e^{2\pi i\alpha}e^{2\pi i\beta}e^{-2\pi i\alpha}e^{-2\pi i\beta}=1\, .
\end{equation}
Thus the Pochhammer contour is closed in twisted homology, even though it is not an ordinary residue contour. With this convention one has
\begin{equation}\label{Beta function}
\oint_{P_u} u^{\alpha-1}(1-u)^{\beta-1}\,du
=
(1-e^{2\pi i\alpha})(1-e^{2\pi i\beta})B(\alpha,\beta),
\end{equation}
where
\begin{equation}
B(\alpha,\beta)
=
\frac{\Gamma(\alpha)\Gamma(\beta)}
{\Gamma(\alpha+\beta)}
\end{equation}
is understood by analytic continuation.

This distinction is important. As an element of the fundamental group of the twice-punctured plane, the Pochhammer contour is a genuine commutator. In ordinary homology, however, it is trivial, since its net winding number around each branch point vanishes. Its nontriviality is visible only after keeping track of the monodromy of the multivalued integrand.

This is precisely what is needed here. If Eq. \eqref{conformally coupled scalar} is to be interpreted as a generalized Penrose transform, it should pair a cohomology class with a closed cycle. This pairing encodes the choice of cohomology class representative, that is, the overlap where the integrand is holomorphic, and where the contour resides. Since the representative is now multivalued, the relevant cycles are twisted cycles. A generic loop around a single branch point is not closed in this sense, because it returns to a different branch.  The Pochhammer contour is closed precisely because the total monodromy \eqref{monodromy} is trivial. Consequently, the resulting period is invariant under deformations of the contour which do not cross singularities.

For the contour appearing below, the relevant punctured curve is
\begin{equation}
\mathbb{CP}^1\setminus\{0,1,\infty\}.
\end{equation}
The corresponding twisted cohomology group is one-dimensional \cite{Matsumoto:2022, Pokraka:2025zlh}. Hence, up to an overall normalization, there is a unique independent twisted period. This matches the fact that the scalar two-point function is fixed up to normalization.

\subsection{Evaluation of Eq. \eqref{eq:Jz-integral}} 
We now evaluate Eq. \eqref{eq:Jz-integral}. Put the two finite branch points at $0$ and $1$ by setting
$z=-\frac{1}{b}+\left(a+\frac{1}{b}\right)u$.
Then, with these change of coordinates we have
\begin{equation}
J_z(a,b)
=
b^{-5/2}\left(a+\frac{1}{b}\right)^{-1}
\oint_{P_u}
u^{-5/2}(1-u)^{1/2}
\left[
\log\left(a+\frac{1}{b}\right)+\log(1-u)
\right]du .
\end{equation}
The term proportional to $\log(a+1/b)$ vanishes and the remaining term is obtained by differentiating the Pochhammer beta integral with respect to $\beta$:
\begin{equation}
\oint_{P_u}
u^{-5/2}(1-u)^{1/2}\log(1-u)\,du
=
\left.
\frac{\partial}{\partial\beta}
\left[
(1-e^{2\pi i\alpha})(1-e^{2\pi i\beta})B(\alpha,\beta)
\right]
\right|_{\alpha=-3/2,\;\beta=3/2}.
\end{equation}
At $\alpha=-3/2$ and $\beta=3/2$, the beta function itself vanishes and hence the derivative of the monodromy prefactor does not contribute while\footnote{This is most easily computed by taking $\beta=\frac{3}{2}+ \epsilon$ and taking the limit of $\epsilon \rightarrow 0$ at the end} 
\begin{equation}
\left.
\frac{\partial B}{\partial\beta}
\right|_{\alpha=-3/2,\;\beta=3/2}
=
\Gamma\left(-\frac{3}{2}\right)
\Gamma\left(\frac{3}{2}\right)
=
\frac{2\pi}{3}.
\end{equation}
It follows that
\begin{equation}
\oint_{P_u}
u^{-5/2}(1-u)^{1/2}\log(1-u)\,du
=
\frac{8\pi}{3}.
\end{equation} 
Thus
\begin{equation}
J_z(a,b)
=
\frac{8\pi}{3}\,
\frac{b^{-3/2}}{1+ab}.
\end{equation}


\bibliography{refs.bib}
\bibliographystyle{JHEP}

\end{document}